\documentclass[12pt]{article}
\usepackage[utf8]{inputenc}%
\usepackage[T1]{fontenc}
\usepackage{amsmath}%
\usepackage{amsfonts}%
\usepackage{amsthm}%
\usepackage{pdflscape}
\usepackage{longtable}
\usepackage{booktabs}
\usepackage{tabularx}
\usepackage{caption}
\usepackage{adjustbox}
\usepackage{subcaption}
\usepackage{multirow}
\usepackage{graphicx}

\theoremstyle{definition}

\theoremstyle{remark}

\usepackage[margin=1.2in]{geometry}
\usepackage[onehalfspacing]{setspace}
\usepackage{xurl}
\usepackage{mathtools}

\usepackage{bbm} 

\usepackage{comment}
\usepackage[title]{appendix}
\usepackage{etoolbox} 
\AtBeginEnvironment{appendices}{\crefalias{section}{appendix}}

\usepackage[
bibencoding=utf8,
maxbibnames=4, 
minbibnames=1, 
maxcitenames=3, 
mincitenames=1, 
backend=biber,%
sortlocale=en_US,%
style=authoryear,
uniquename=false,%
url=true,%
sortcites=false,
doi=true,%
eprint=true%
]{biblatex}
\usepackage{changepage}

\usepackage{xcolor}%
\definecolor{webbrown}{rgb}{.6,0,0}%
\usepackage{hyperref} %
\hypersetup{%
  breaklinks = true,%
  colorlinks = true,%
  anchorcolor = webbrown,%
  citecolor = webbrown,%
  filecolor = webbrown,%
  linkcolor = webbrown,%
  menucolor = webbrown,%
  urlcolor= webbrown,%
  citebordercolor= 1 0 0,%
  menubordercolor=1 0 0,%
  urlbordercolor=1 0 0,%
  runbordercolor=1 0 0,%
  pdftitle={Tracking the Credibility Revolution across Fields},
  pdfauthor={Goldsmith-Pinkham}
}

\usepackage{cleveref}
\crefname{appsec}{appendix}{appendices}
\crefname{appsubsec}{appendix}{appendices}
\crefname{assumption}{assumption}{assumptions}

\usepackage[nolist]{acronym}

\usepackage{accents}

\newcolumntype{Y}{>{\centering\arraybackslash}X}

\usepackage{float}
\title{Tracking the Credibility Revolution across Fields\thanks{\noindent Goldsmith-Pinkham:
    \href{mailto:paul.goldsmith-pinkham@yale.edu}{paul.goldsmith-pinkham@yale.edu}}}%
\author{
  Paul Goldsmith-Pinkham\\
  Yale University and NBER
}
\date{May 2024}

\begin{document}
\maketitle
\thispagestyle{empty} 
\setcounter{page}{0}
\begin{abstract}
  This paper updates \textcite{currie2020technology} by examining the credibility revolution across fields, including finance and macroeconomics, using NBER working papers up to May 2024. While the growth in terms related to identification and research designs have continued, finance and macroeconomics have lagged behind applied micro. Difference-in-differences and regression discontinuity designs have risen since 2002, but the growth in difference-in-difference has been larger, more persistent, and more ubiquitous. In contrast, instrumental variables have stayed flat over this period. Finance and macro, particularly corporate finance, has experienced significant growth in mentions of experimental and quasi-experimental methods and identification over this time period, but a large component of the credibility revolution in finance is due to difference-in-differences. Bartik and shift-share instruments have grown across all fields, with the most pronounced growth in international trade and investment, economic history, and labor studies. Synthetic control has not seen continued growth, and has fallen since 2020. 
\end{abstract}



\clearpage

The credibility revolution in economics has been one of the defining trends of the last two decades \parencite{angrist2010credibility}. Leveraging a wealth of new data and a focus on transparent, credible research designs, economists have generated profound insights into an array of pressing questions, from the determinants of economic growth to the impacts of social and educational policies. But to what extent has the credibility revolution permeated different fields of economics? Are the trends identified in the early days of the movement by \textcite{angrist2010credibility} still continuing apace? And are different empirical methods being adopted evenly, or are some techniques leading the charge while others languish?

In this paper, I take up these questions by building on the innovative approach of \textcite{currie2020technology}. Using natural language processing methods, I analyze the text of over 32,000 National Bureau of Economic Research (NBER) working papers to identify the frequency of phrases related to different empirical techniques. By extending the sample period of Currie et al. (2020) to 2024 and including papers from all NBER programs, I assess the latest state of the credibility revolution across fields, including finance and macroeconomics, which were omitted from the original analysis.

The results show that the aggregate trends identified by Currie et al. (2020) are still advancing, with the use of experimental and quasi-experimental methods continuing to rise through 2024. However, there is significant heterogeneity across fields. While applied microeconomics has embraced empirical techniques that emphasize research design, such as difference-in-differences, event studies, and randomized trials, finance and macroeconomics lag behind. 

Within finance, corporate finance has seen robust growth in a subset of these tools, particularly difference-in-differences designs, but the use of instrumental variables, regression discontinuity, and experimental methods is a much smaller relative share. The adoption of popular tools like Bartik and shift-share instruments is also highly uneven across fields, with rapid growth in applied micro areas like labor, trade, and economic history, while other tools like synthetic controls appear to have already peaked in their popularity.

These findings provide an important check on the sometimes triumphalist narrative of the credibility revolution rapidly sweeping across economics. They suggest a more nuanced picture, with the frontier of empirical work using credible, transparent research designs still centered in applied microeconomics, and other fields making strides but at an uneven pace. The analysis also highlights how a single method, in this case difference-in-differences, can dominate the rise of empirical work in a field like finance and macroeconomics, when other techniques are not taken up in parallel.

\section{Data and Methods}
To construct measures of the use of empirical methods over time, I follow the approach in \textcite{currie2020technology}. This involves using the text of papers and looking for sets of keywords using regular expressions that capture the spirit of the credibility revolution (e.g. ``threats to identification'' or ``identification strategy'').\footnote{See the Appendix for the full set of words. I follow the same method as \textcite{currie2020technology}.}. I replicate \textcite{currie2020technology}'s data collection process for the NBER working paper series, starting from working paper 1000 to 32,436. Unlike \textcite{currie2020technology}, I do not exclusively focus on papers under the ``applied micro'' heading, but instead include all papers in the NBER working paper series. After text processing and cleaning, I am left with a sample of 28,397 papers from 1982 to 2024. 

\begin{figure}[thbp]
  \centering
  \begin{subfigure}[t]{0.5\textwidth}
    \includegraphics*[width=\linewidth]{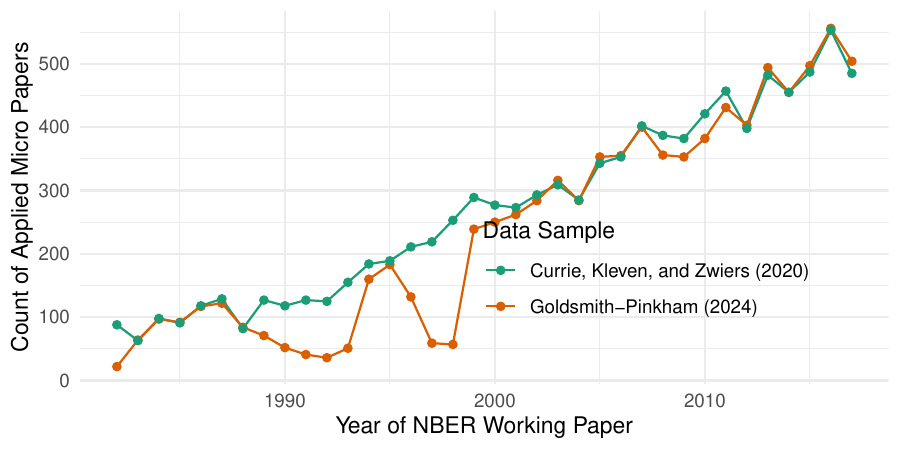}
    \caption{Comparison of sample size to \textcite{currie2020technology} in  ``applied micro''}
    \label{fig:comparison}
  \end{subfigure}%
  ~ 
  \begin{subfigure}[t]{0.5\textwidth}
    \includegraphics*[width=\linewidth]{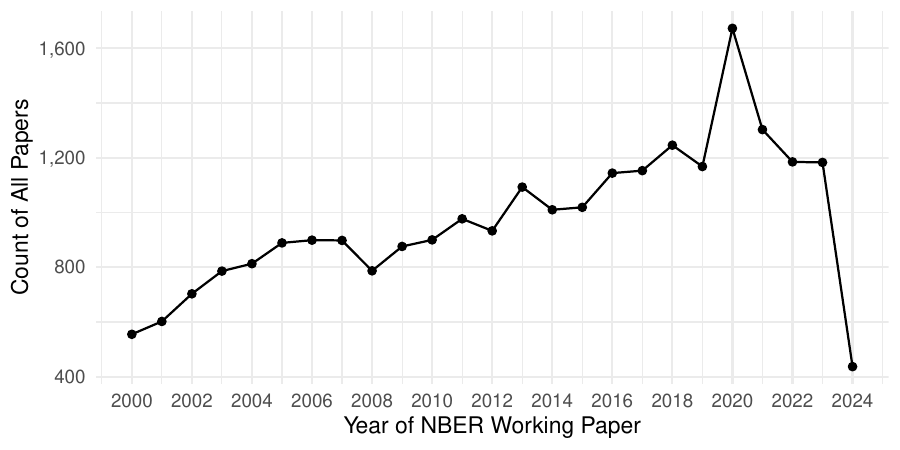}
    \caption{Total papers in final sample over time}
    \label{fig:sample_over_time}
  \end{subfigure}
  \caption{NBER Working Paper Counts over Time. Data for \textcite{currie2020technology} is measured in Appendix Figure B.I. in their paper. My sample ends in May 2024.}
\end{figure}

\begin{figure}[thbp]
  \centering
  \begin{subfigure}[t]{0.5\textwidth}
    \includegraphics*[width=\linewidth]{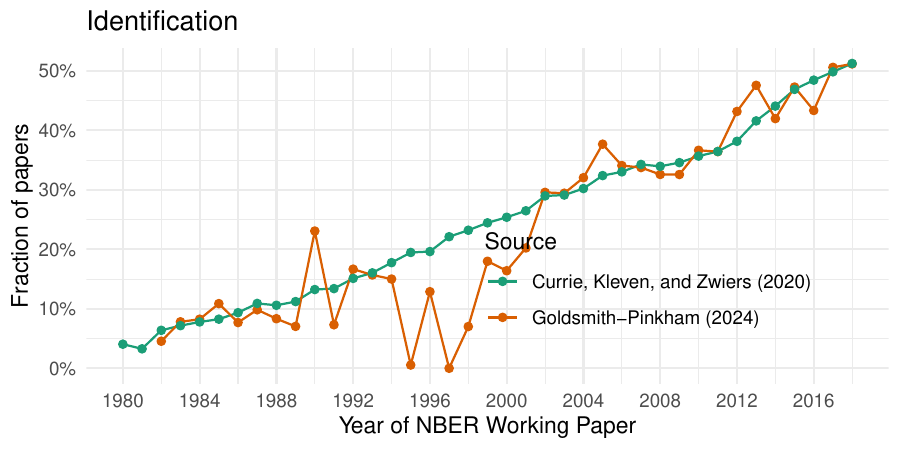}
    \caption{Comparison of identification measure to \textcite{currie2020technology} in  ``applied micro''}
    \label{fig:identification_comparison}
  \end{subfigure}%
  ~ 
  \begin{subfigure}[t]{0.5\textwidth}
    \includegraphics*[width=\linewidth]{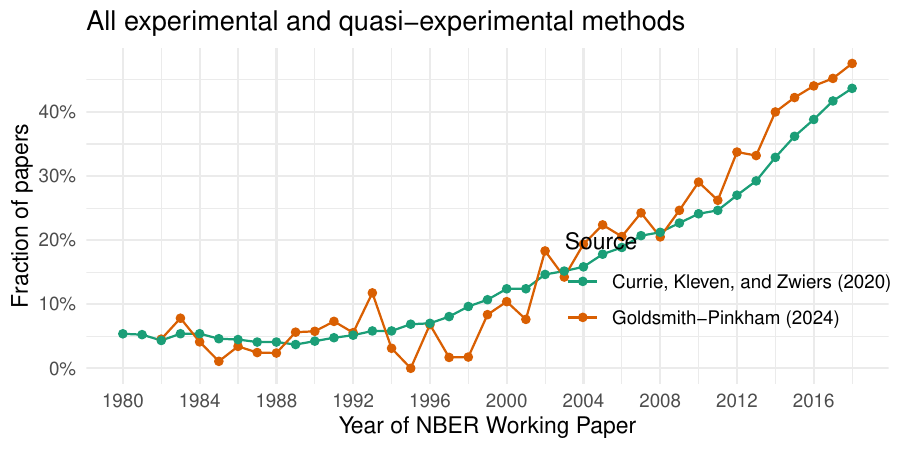}
    \caption{Comparison of all experimental and quasi-experimental measure to \textcite{currie2020technology} in  ``applied micro''}
    \label{fig:all_experimental_comparison}
  \end{subfigure}
  \caption{Validation of measurement with \textcite{currie2020technology}. Data for \textcite{currie2020technology} is taken from Figure 2 Panel A and B. I plot the raw measure, while the \textcite{currie2020technology} measure is a rolling five-year mean.}
  \label{fig:validation2}
\end{figure}

As discussed in \textcite{currie2020technology}'s replication package \parencite{currie2020data}, the data cleaning steps for the PDFs can cause errors, particularly in the PDF-to-text conversion. As a validation exercise, I compare the number of papers in my sample over time in the ``applied micro'' setting to \textcite{currie2020technology} in \Cref{fig:comparison}. My sample has many more gaps in the 1990s, due to data processing errors for PDFs in that time period, but coverage appears very close in the early 1980s, and from 1999 onwards. In \Cref{fig:validation2}, I compare two headline estimates from \textcite{currie2020technology} to my estimates -- the rise in the fraction of papers making explicit reference to identification (``identification'') and the fraction of papers making reference to  randomized controlled trials (RCTs), lab experiments, difference-in-differences, regression discontinuity, event studies, or bunching (``All experimental and quasi-experimental methods''). In both settings, my estimates track reasonably well, except for during the late 1990s period. As a result, in all results going forward, I focus on the period of 2000 onwards for the remainder of the analysis, leaving a sample of 24,702 papers. I plot the sample over time in \Cref{fig:sample_over_time}.

In the NBER working paper series, papers can be submitted to different programs (there are fourteen in total). A single paper may be submitted to multiple programs, and 55\% of papers are submitted to more than one.\footnote{45\% have one program, 32\% have two programs, 15\% have three programs, 5\% have four programs, and 2\% have five programs.} I report the breakdown by program in \Cref{tab:program_counts}. The most common programs are Economic Fluctuations and Growth (a macroeconomics research program),  Public Economics (an ``applied micro'' research program, as classified by \textcite{currie2020technology}) and Labor Studies, (also ``applied micro''). Development Economics is smaller in part because it only began as a program in 2012.

\begin{table}[thbp]
  \footnotesize
  \begin{center}
  
\begin{tabular}{llr}
\toprule
NBER Program & Field & Number of Papers\\
\midrule
Asset Pricing & Finance & 2,739\\
Children & Applied Micro & 1,651\\
Corporate Finance & Finance & 2,502\\
Development Economics & Applied Micro & 1,070\\
Development of the American Economy & Macro/Others & 1,398\\
\addlinespace
Economic Fluctuations and Growth & Macro/Others & 3,952\\
Economics of Aging & Applied Micro & 1,185\\
Economics of Education & Applied Micro & 1,584\\
Economics of Health & Applied Micro & 2,636\\
Environment and Energy Economics & Applied Micro & 1,143\\
\addlinespace
Industrial Organization & Applied Micro & 1,023\\
International Finance and Macroeconomics & Macro/Others & 2,167\\
International Trade and Investment & Applied Micro & 1,179\\
Labor Studies & Applied Micro & 3,387\\
Law and Economics & Macro/Others & 995\\
\addlinespace
Monetary Economics & Macro/Others & 1,732\\
Political Economy & Applied Micro & 857\\
Productivity, Innovation, and Entrepreneurship & Macro/Others & 2,066\\
Public Economics & Applied Micro & 3,413\\
\bottomrule
\end{tabular}

  \caption{NBER Working Paper Series counts by program}
  \label{tab:program_counts}
\end{center}
\end{table}

In order to provide a simple comparison across programs, I extend \textcite{currie2020technology}'s classification of applied micro. I define ``finance'' as Asset Pricing and Corporate Finance, while ``macro'' or ``macro/other'' includes the remainder of programs: Development of the American Economy (the economic history research group), Economic Fluctuations and Growth, International Finance and Macroeconomics, Law and Economics, Monetary Economics, and  Productivity Innovation and Entrepreneurship. These fields are defined in \Cref{tab:field_groups}. \textcite{currie2020technology} go one step further and define applied micro as papers that \emph{solely} have applied micro programs listed for them, making applied micro a paper-specific label. To simplify the analysis, I use non-exclusive labels -- if a paper is in both finance and applied micro, it is counted in both categories. The amount of overlap is non-trivial, but not extreme. In \Cref{tab:field_groups}, I report the overlap across fields. Roughly 44 percent of papers are excusively applied micro, seven percent in excusively finance, 19 percent in just macro/other, and then a scattering across other pairings. The most common cross field pairings are applied micro and macro/other (19\%), and finance and macro/other (6\%).

\begin{table}
  \footnotesize
  \begin{center}
  
\begin{tabular}{lrr}
\toprule
Field & Number of Papers & Share of Papers\\
\midrule
Applied Micro & 10,654 & 0.44\\
Applied Micro, Finance & 535 & 0.02\\
Applied Micro, Finance, Macro/Others & 657 & 0.03\\
Applied Micro, Macro/Others & 4,521 & 0.19\\
Finance & 1,633 & 0.07\\
\addlinespace
Finance, Macro/Others & 1,550 & 0.06\\
Macro/Others & 4,682 & 0.19\\
\bottomrule
\end{tabular}

  \caption{Breakdown of papers by field groupings}
  \label{tab:field_groups}
  \end{center}
\end{table} 

In the results that follow, any field or program-specific results are not mutually exclusive. A paper contributes equally to each program that it is submitted to, and I interpret the results accordingly.

\section{Results}
\subsection{Overall trends}
\begin{figure}[h]
  \centering
  \begin{subfigure}[t]{0.5\textwidth}
    \includegraphics*[width=\linewidth]{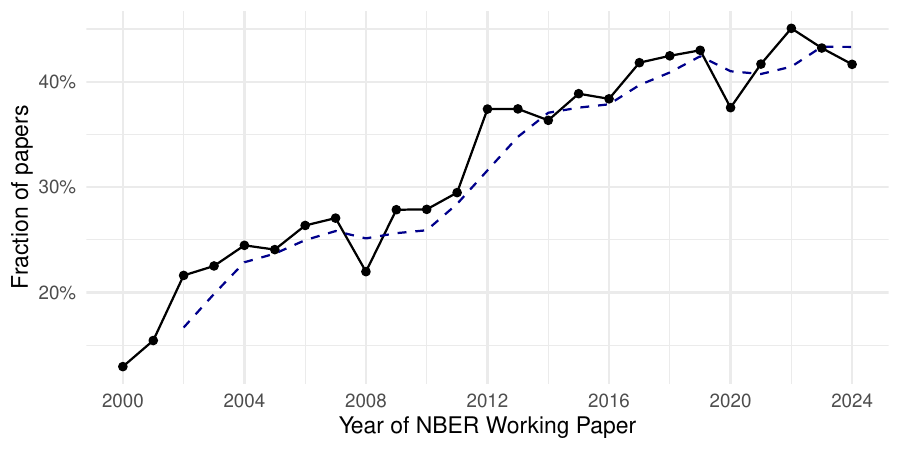}
    \caption{Identification}
    \label{fig:identification}
  \end{subfigure}%
  ~ 
  \begin{subfigure}[t]{0.5\textwidth}
    \includegraphics*[width=\linewidth]{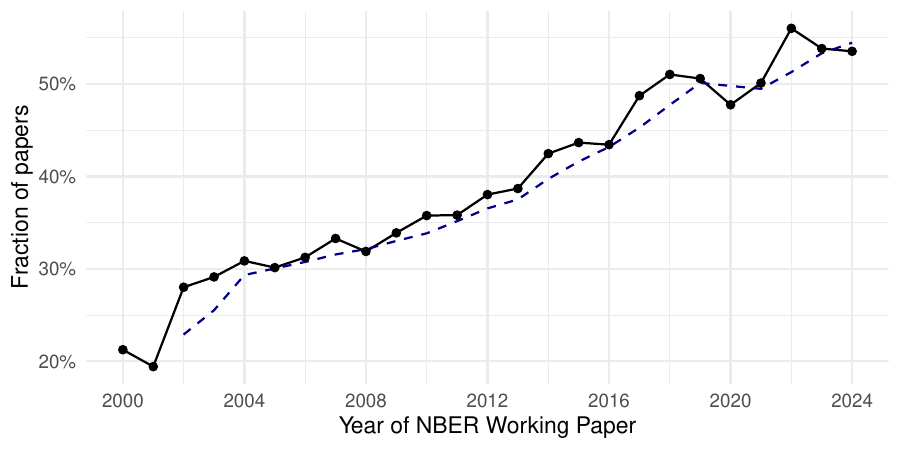}
    \caption{All experimental and quasi-experimental methods}
    \label{fig:all_experimental}
  \end{subfigure}
  \begin{subfigure}[t]{0.5\textwidth}
    \includegraphics*[width=\linewidth]{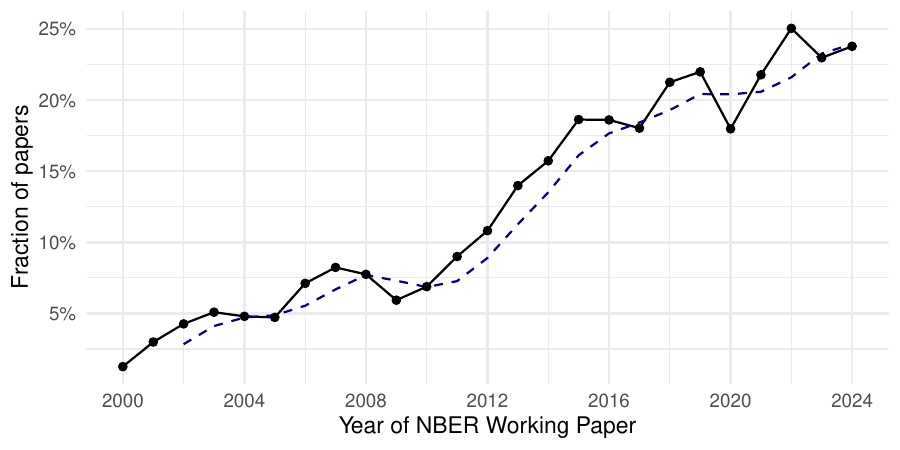}
    \caption{Administrative data}
    \label{fig:admin_data}
  \end{subfigure}%
  ~ 
  \begin{subfigure}[t]{0.5\textwidth}
    \includegraphics*[width=\linewidth]{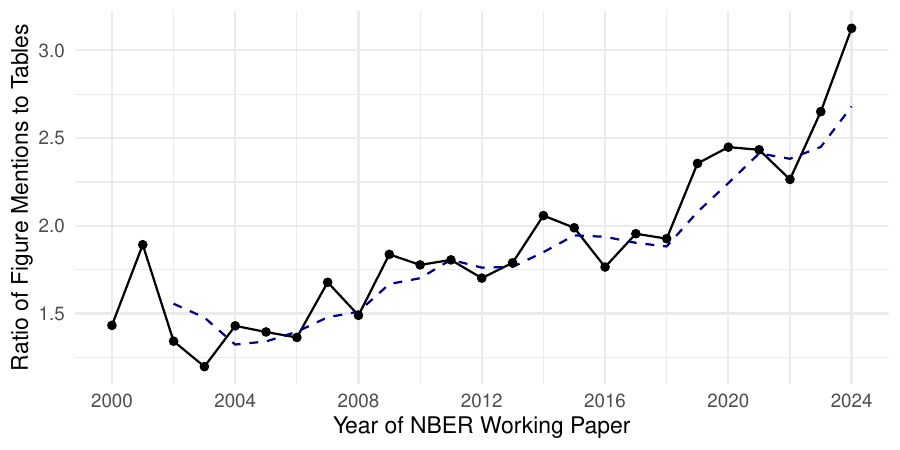}
    \caption{Graphical revolution}
    \label{fig:graph_rev}
  \end{subfigure}
  \caption{This figure updates the main Figure 2 from \textcite{currie2020technology} using all papers (not just applied micro) and extended until May 2024. Figure (a) reports the share of papers that mention identification strategies or concerns. Figure (b) reports the share of papers that mention any experimental or quasi-experimental method (this includes diff-in-diff, event studies, regression discontinuity, randomized control trials, lab experiments, bunching designs, and instrumental variables). Note that the original \textcite{currie2020technology} measure does not include instrumental variables. Figure (c) reports the share of papers that mention administrative data. Figure (d) reports the average number of figure mentions relative to table mentions.}
  \label{fig:extend}
\end{figure}

I begin by presenting the updated overall trends over the sample period for all papers. In \Cref{fig:extend}, I present the updated version of \textcite{currie2020technology}'s Figure 2.\footnote{In \textcite{currie2020technology}, they use a five year moving average for their results, whereas I present either the raw underlying data, or a two-year moving average, or both.} For each graph, the solid line reflects the raw data and the dashed line is the two-year moving average.

Almost all trends are similar to \textcite{currie2020technology}'s sample, and continued in the same direction. In \Cref{fig:identification}, the share of papers that make explicit mention to identification has gone up overall, but has flattened since 2016 at around 40\%.\footnote{The keywords used are available in the Appendix, as well as the Appendix of \textcite{currie2020technology}. To give a sense of what this looks for, matches for identification would flag things like ``identification assumption'' and ``causal identification.''} In \Cref{fig:all_experimental}, the share of papers that make reference to any experimental or quasi-experimental method has continued to rise, even after 2016. Growth in administrative data (\Cref{fig:admin_data}) and the graphical revolution (\Cref{fig:graph_rev}, which calculates the share of figure mentions relative to table mentions) have also continued.

\subsection{Comparison across fields}
\begin{figure}[thbp]
  \centering
  \begin{subfigure}[t]{0.5\textwidth}
    \includegraphics*[width=\linewidth]{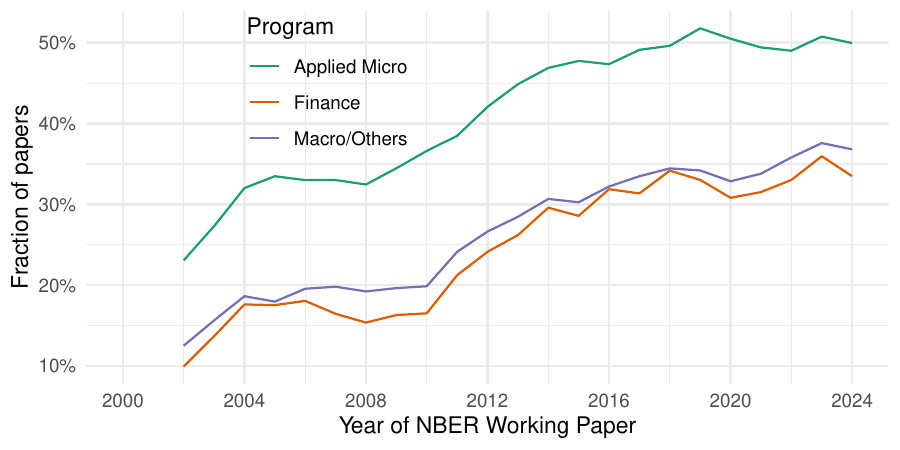}
    \caption{Identification}
    \label{fig:identification_fields}
  \end{subfigure}%
  ~ 
  \begin{subfigure}[t]{0.5\textwidth}
    \includegraphics*[width=\linewidth]{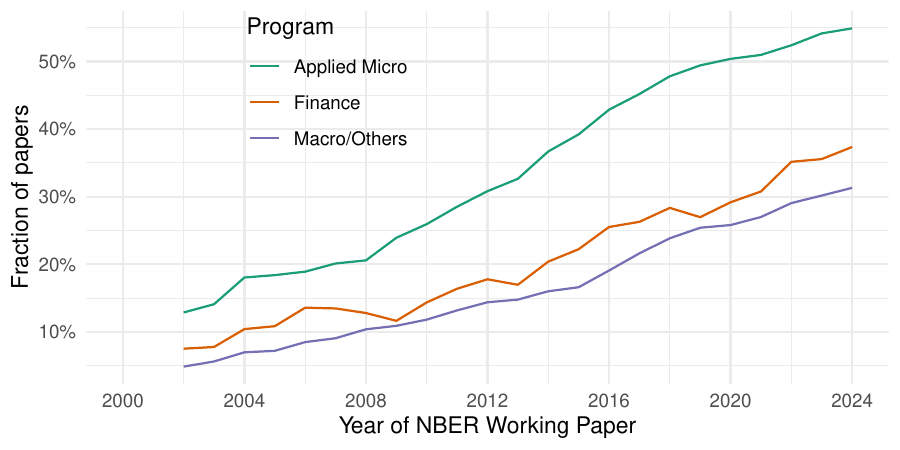}
    \caption{All experimental and quasi-experimental methods}
    \label{fig:all_experimental_comparison_fields}
  \end{subfigure}
  \begin{subfigure}[t]{0.5\textwidth}
    \includegraphics*[width=\linewidth]{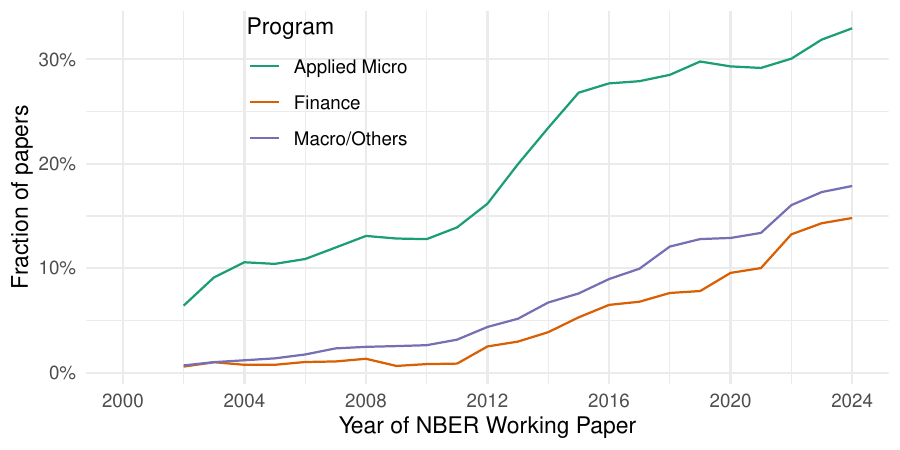}
    \caption{Administrative data}
    \label{fig:admin_data_fields}
  \end{subfigure}%
  ~ 
  \begin{subfigure}[t]{0.5\textwidth}
    \includegraphics*[width=\linewidth]{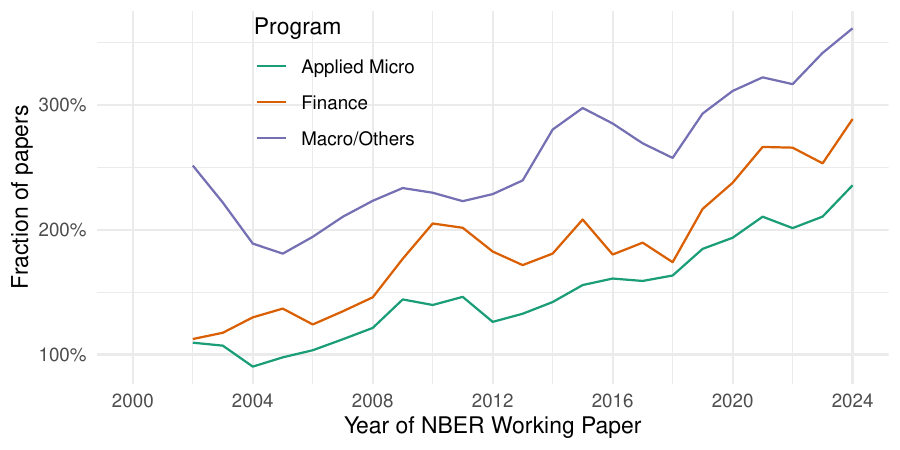}
    \caption{Graphical revolution}
    \label{fig:graph_rev_fields}
  \end{subfigure}
  \caption{ This figure splits \Cref{fig:extend} into three overlapping sets of  papers: papers submitted to an applied micro group, papers submitted to a finance group,and papers submitted to a macro (or other) group. Figure (a) reports the share of papers that mention identification strategies or concerns. Figure (b) reports the share of papers that mention any experimental or quasi-experimental method (this includes diff-in-diff, event studies, regression discontinuity, randomized control trials, lab experiments, bunching designs, and instruemntal variables). Note that the original \textcite{currie2020technology} measure does not include instrumental variables. Figure (c) reports the share of papers that mention administrative data. Figure (d) reports the average number of figure mentions relative to table mentions. See \Cref{tab:field_groups} for the breakdown of fields, and the Appendix for definitions on keywords.}
  \label{fig:extend_byfield}
\end{figure}

I next turn to the comparison across fields, using the breakdown defined in \Cref{tab:field_groups}, and replicate \Cref{fig:extend} split by field in \Cref{fig:extend_byfield}.\footnote{I report the two-year moving average to simplify the figures.} There are a few distinctive patterns that stand out. For mentions of identification, experimental and quasi-experimental methods, and admin data, applied micro is significantly higher than both finance and macro/other. In mentions of identification, applied micro has roughly plateaued as of 2017 at 50\%, but remains 15\% higher than finance and macro/others as of 2024. In share of papers with mentions of methods in \Cref{fig:all_experimental_comparison_fields}, applied micro is at roughly 55\% as of 2024, while finance has risen to only 38\% and macro/other is at a little over 30\%. The graphical revolution in \Cref{fig:graph_rev_fields} is reversed -- macro/other has the highest share of figures relative to tables, followed by finance, and then applied micro.

Since the credibility revolution has grown and permeated the entire economics profession, one useful summary of these results is to examine the state of finance and macro/other \emph{now} relative to applied micro in the past. In terms of mentions of identification, finance and macro/other are roughly where applied micro was between 2008 and 2010. In terms of experimental and quasi-experimental methods, finance and macro/other are roughly where applied micro was between 2012 and 2014. In terms of admin data, finance and macro/other are roughly where applied micro was in 2013.

\begin{figure}[tbhp]
  \centering
  \begin{subfigure}[t]{0.5\textwidth}
    \includegraphics*[width=\linewidth]{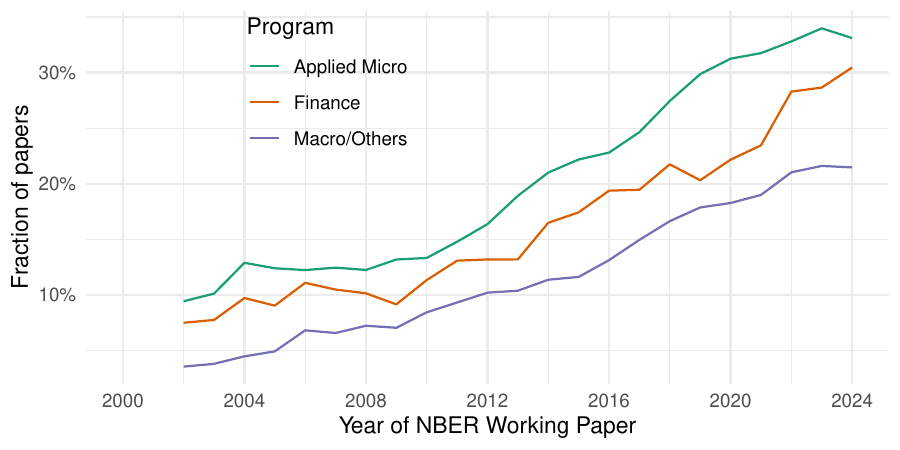}
    \caption{Difference-in-differences}
    \label{fig:did_by_field}
  \end{subfigure}%
  ~ 
  \begin{subfigure}[t]{0.5\textwidth}
    \includegraphics*[width=\linewidth]{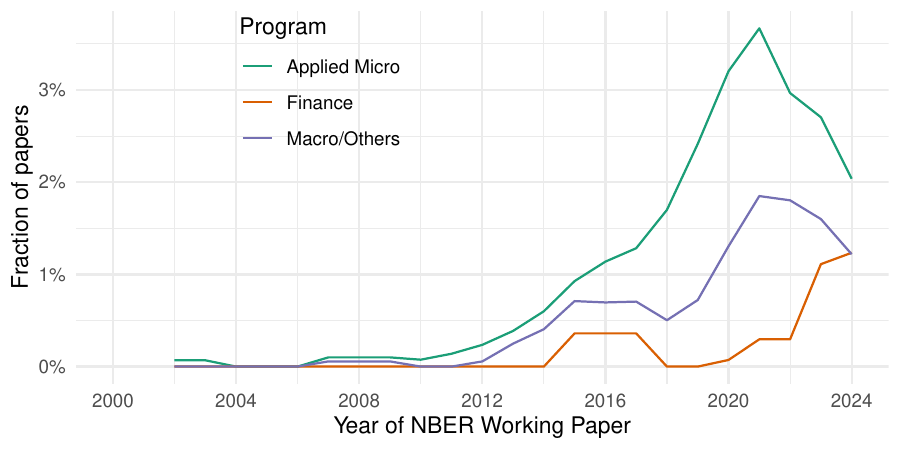}
    \caption{Synthetic controls}
    \label{fig:synth_by_field}
  \end{subfigure}
  
  \caption{Panel (a) reports the share of papers that mention difference-in-differences or event studies. Figure (b) reports the share of papers that mention synthetic controls (this includes both synthetic difference-in-differences and synthetic control methods). See \Cref{tab:field_groups} for the breakdown of fields, and the Appendix for definitions on keywords.}
  \label{fig:did_synth_fields}
\end{figure}

I next turn to specific identification methods across fields. In \Cref{fig:did_by_field}, I plot the growth in difference-in-differences over time across the three fields. This includes mentions of both difference-in-differences and event studies. For all three fields, there has been significant growth in difference-in-differences, with applied micro leading the way. However, finance is close behind, in part due to the term ``event study,'' which captures many financial event studies (that are distinct in their design from traditional difference-in-differences). 

I next examine the growth in synthetic controls. Notably, in Appendix Figure A.V in \textcite{currie2020technology}, synthetic control was experiencing rapid growth as of 2018 among applied micro papers. \Cref{fig:synth_by_field} shows that this growth continued until 2020 but appears to have fallen since. Much of this growth was concentrated in applied micro and macro. This suggests that take-up has slowed, and may have even fallen.

\begin{figure}[thbp]
  \centering
  \begin{subfigure}[t]{0.5\textwidth}
    \includegraphics*[width=\linewidth]{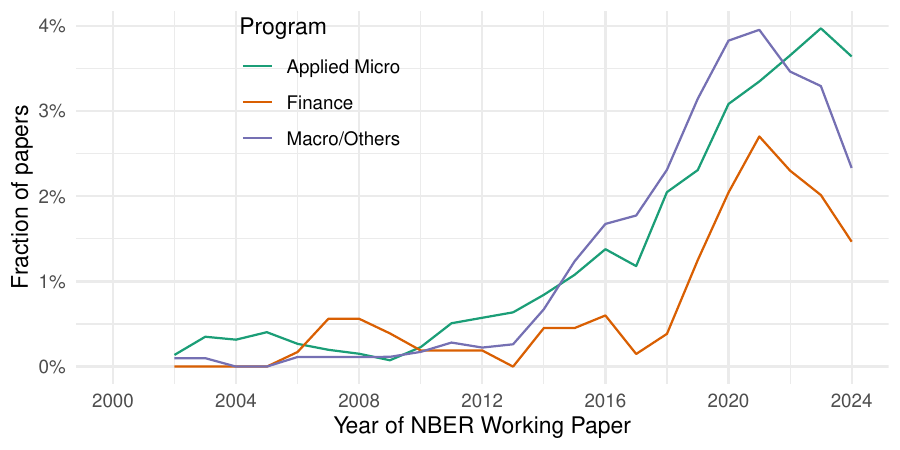}
    \caption{Bartik and shift-share instruments}
    \label{fig:bartik_by_field}
  \end{subfigure}%
  ~ 
  \begin{subfigure}[t]{0.5\textwidth}
    \includegraphics*[width=\linewidth]{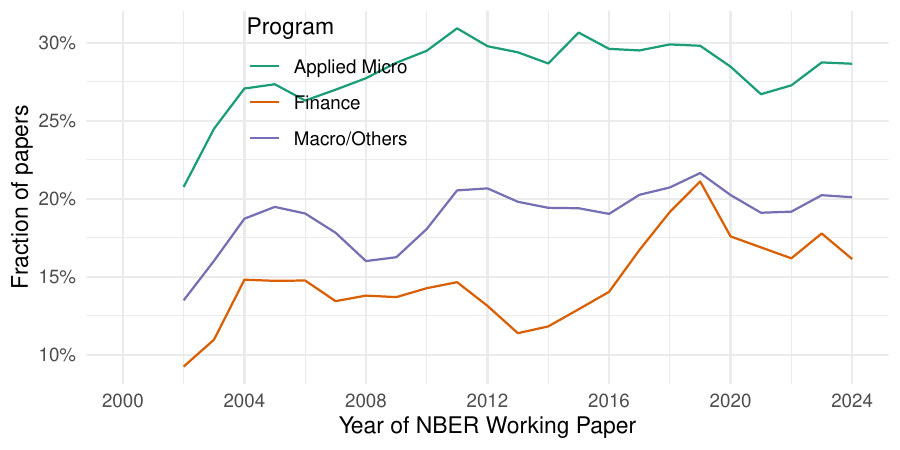}
    \caption{Instrumental variables}
    \label{fig:iv_by_field}
  \end{subfigure}
  
  \caption{Panel (a) reports the share of papers that mention Bartik or shift-share instruments. Figure (b) reports the share of papers that mention instrumental variables. See \Cref{tab:field_groups} for the breakdown of fields, and the Appendix for definitions on keywords.}
  \label{fig:bartik_iv_fields}
\end{figure}

Next, in a slightly self-indungent fashion \parencite{goldsmith2020bartik}, I examine the rise of Bartik and shift-share instruments in \Cref{fig:bartik_by_field}. Since 2013, this method has grown rapidly across all fields, but with some fall off in macro/others and finance after 2021. Nonetheless, almost 2-4\% of all papers in 2024 mention Bartik or shift-share. To put this in context, I plot the share mentioning instrumental variables at all in \Cref{fig:iv_by_field}. This share has stayed relatively constant over time, with roughly 30\% of applied micro papers, 20\% of macro/others and 15\% of finance papers. Hence, a rough back-of-the-envelope calculation would suggest that, given 2\% of finance papers and 4\% of applied micro papers mention Bartik, 13\% of instrument approaches in finance and applied micro are Bartik or shift-share. 

\begin{figure}[thbp]
  \centering
  \begin{subfigure}[t]{0.5\textwidth}
    \includegraphics*[width=\linewidth]{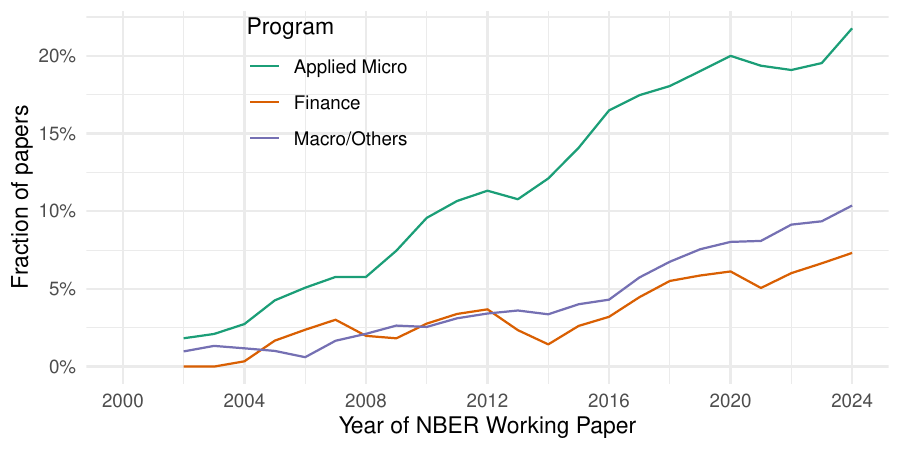}
    \caption{Experiments}
    \label{fig:experiments}
  \end{subfigure}%
  ~ 
  \begin{subfigure}[t]{0.5\textwidth}
    \includegraphics*[width=\linewidth]{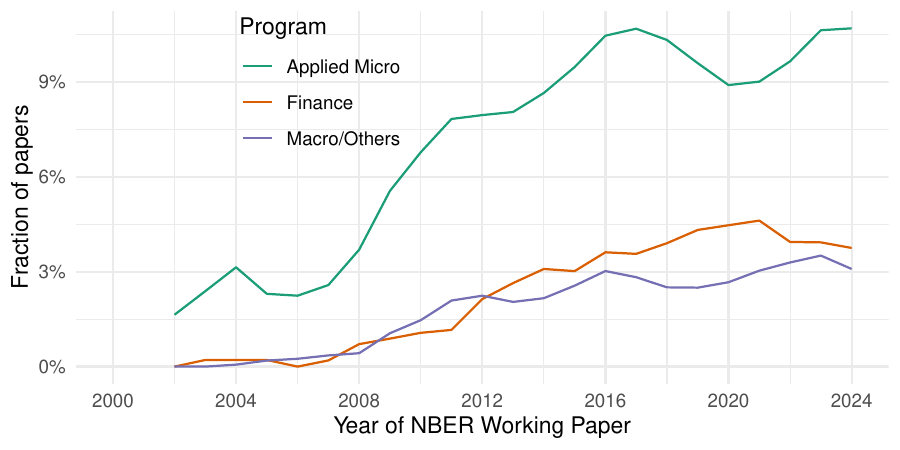}
    \caption{Regression discontinuity}
    \label{fig:rd}
  \end{subfigure}
  
  \caption{Panel (a) reports the share of papers that mention randomized control trials or lab experiments. Figure (b) reports the share of papers that mention regression discontinuity designs. See \Cref{tab:field_groups} for the breakdown of fields, and the Appendix for definitions on keywords.}
  \label{fig:did}
\end{figure}

Finally, I examine the use of experiments (randomized control trials) and regression discontinuity designs across fields. In \Cref{fig:experiments}, I plot the share of papers that mention randomized control trials. Here, applied micro is the clear leader, with 20\% of papers mentioning RCTs in 2024. For both macro and finance, this share has grown as well, but less. Strikingly, all three fields had a relatively similar base as of 2003.

In \Cref{fig:rd}, I plot the share of papers that mention regression discontinuity designs. Here, applied micro is roughly 6 percentage points higher than finance and macro/other as of 2024, but for all fields the share flattened in the past 8 years.

\begin{figure}[thbp]
  \centering
  \begin{subfigure}[t]{0.5\textwidth}
    \includegraphics*[width=\linewidth]{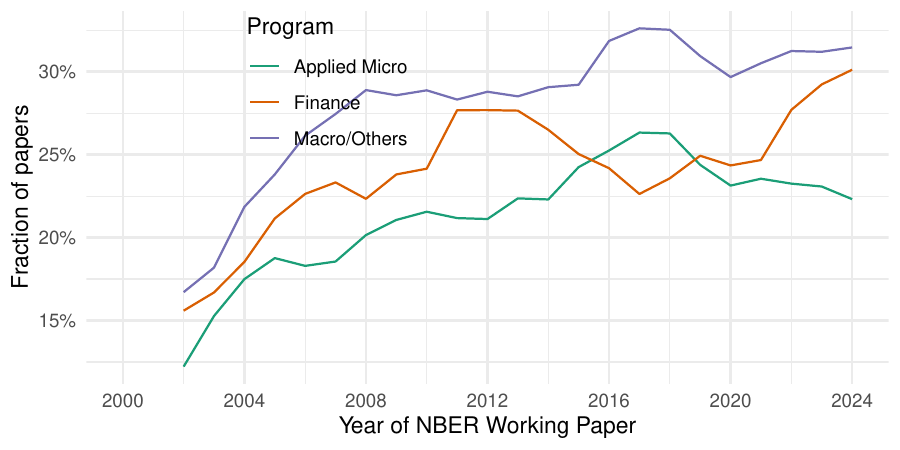}
    \caption{Structural Models}
    \label{fig:structural_by_field}
  \end{subfigure}%
  ~ 
  \begin{subfigure}[t]{0.5\textwidth}
    \includegraphics*[width=\linewidth]{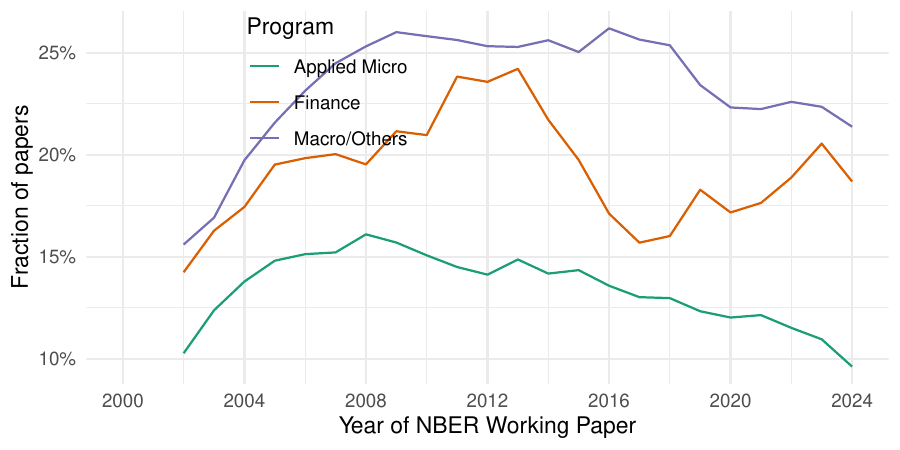}
    \caption{Structural Models without mention of experimental or quasi-experimental methods}
    \label{fig:structural_by_field_no_quasi}
  \end{subfigure}
  
  \caption{Panel (a) reports the share of papers that mention structural model estimation. Figure (b) reports the share of papers that mention structural model estimation and do not mention any form of experimental or non-experimental methods. See \Cref{tab:field_groups} for the breakdown of fields, and the Appendix for definitions on keywords.}
  \label{fig:did}
\end{figure}

One natural question is what types of papers sit in the gap between applied micro and the other fields. Some of this may be pure theory or observational papers. One alternative (already measured in \textcite{currie2020technology}) is to look at the share of papers that mention structural estimation (including words like ``structural estimation'' and ``structural model'' or ``structural general equilibrium model'' or ``GMM''). In \Cref{fig:structural_by_field}, I plot the share of papers that mention structural estimation. Here, macro/others and finance tend to have a 7.5-10 percent higher share of papers, consistent with the idea that these fields may have more structural models. But, it is worth recalling that applied micro includes Industrial Organization. It is also useful to identify the set of papers that do not mention experimental or quasi-experimental methods and do mention structural estimation. I plot this share in \Cref{fig:structural_by_field_no_quasi}. Here, the gap between applied micro and the other fields is larger, with 20\% of finance and macro/other papers used structural estimation as of 2024, and only 10\% in applied micro. This suggests that in applied micro papers using structural models, there is more discussion of additional research designs than in finance or macro. 

\subsection{Breakdown across programs}
What underlying variation drives the trends across fields? As an example, among finance, the types of analyses and approaches for asset pricing and corporate finance are very different, and it is reasonable to assume that these programs may have different levels and trends. This is similarly true for applied micro and macro/other.

\begin{figure}[thbp]
  \centering
  \begin{subfigure}[t]{0.5\textwidth}
    \includegraphics*[width=\linewidth]{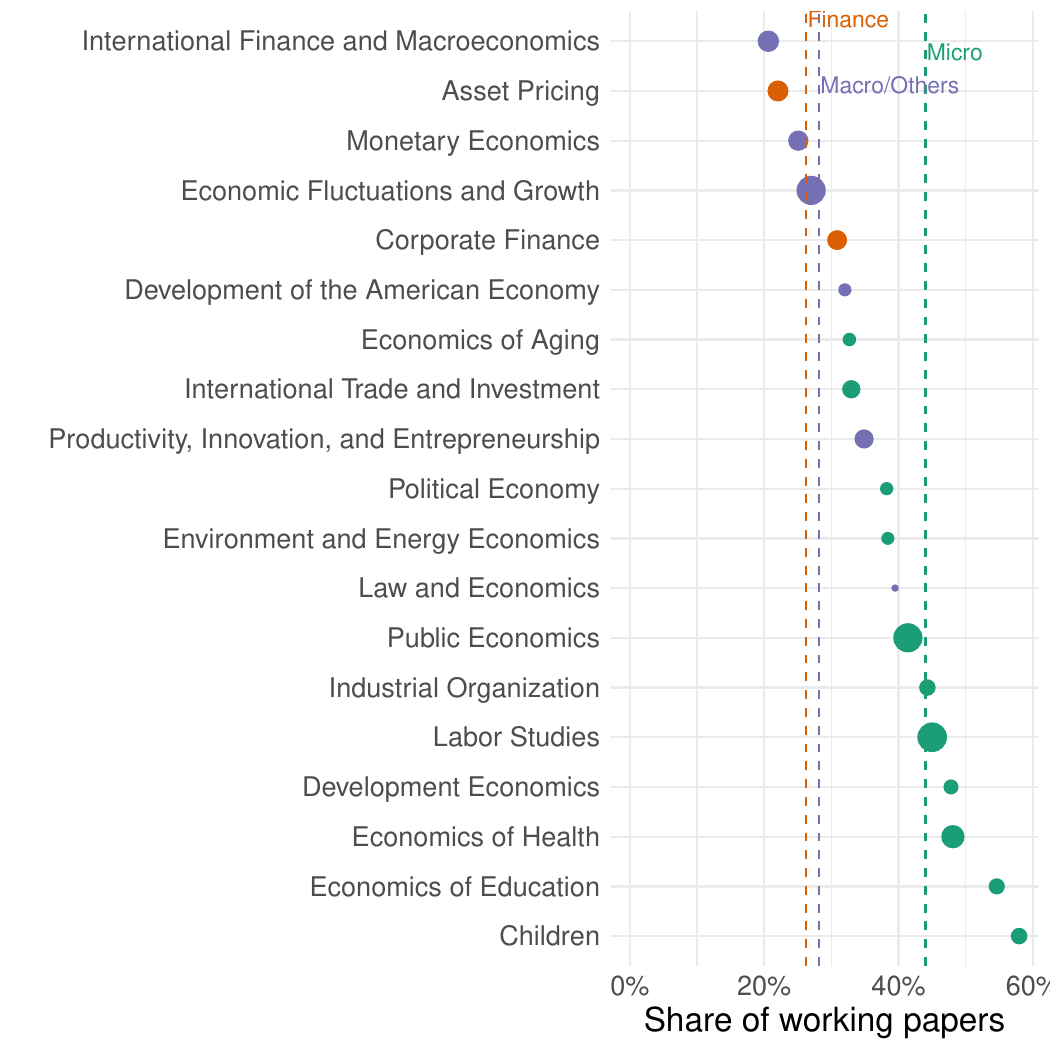}
    \caption{Identification}
    \label{fig:identification_bygroup}
  \end{subfigure}%
  ~ 
  \begin{subfigure}[t]{0.5\textwidth}
    \includegraphics*[width=\linewidth]{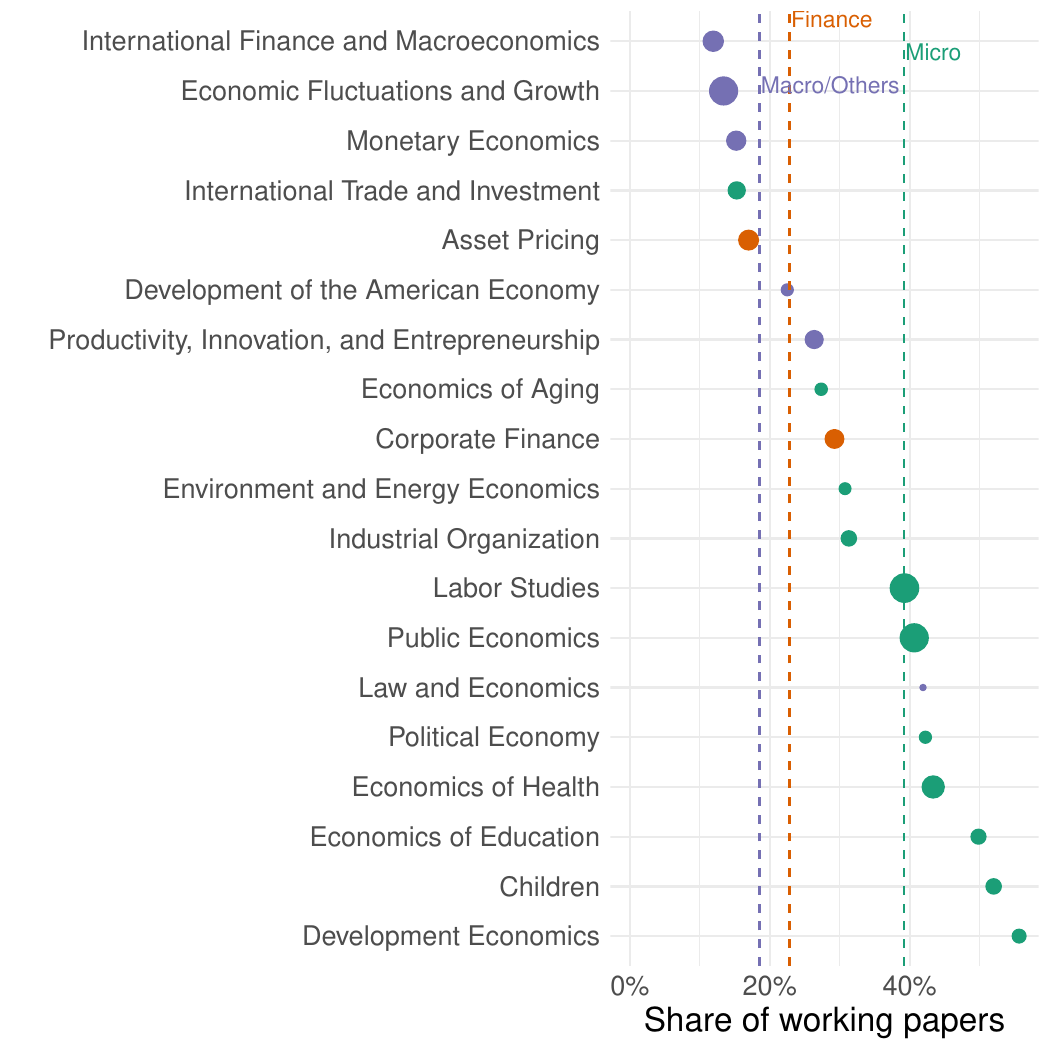}
    \caption{All experimental and quasi-experimental methods}
    \label{fig:all_experimental_comparison}
  \end{subfigure}
  \caption{This Figure splits out by papers into each of the research programs for which a paper can be submitted. The size of each dot reflects the total number of papers in the program. The vertical dotted lines are the average for each field. Papers can be included in more than one research program. Panel (a) reports the share of papers that mention identification strategies or concerns. Panel (b) reports the share of papers that mention any experimental or quasi-experimental method (this includes diff-in-diff, event studies, regression discontinuity, randomized control trials, lab experiments, bunching designs, and instrumental variables).  See \Cref{tab:field_groups} for the breakdown of fields, and the Appendix for definitions on keywords.}
  \label{fig:extend_bygroup}
\end{figure}

In \Cref{fig:extend_bygroup}, I plot the overall share of papers mentioning identification and experimental and quasi-experimental methods across programs. For each figure, the size of the dot reflects the number of papers, the color reflects the program, and the dots are ordered by their relative share. I plot the vertical weighted average for the overall field in the dotted line, corresponding to the overall average from the previous field graphs. 

Here, despite heterogeneity within fields, the breakdown across programs is relatively consistent. For example, in \Cref{fig:identification_bygroup}, the share of papers mentioning identification  in applied micro programs is higher than all finance and macro/other programs, with the exception of Productivity, Innovation, and Entrepreneurship, and Law and Economics. However, there is a large gap between Asset Pricing and Corporate Finance, suggesting that the rise in identification in finance is driven by Corporate Finance. In \Cref{fig:all_experimental_comparison}, the share of papers mentioning experimental and quasi-experimental methods is higher in applied micro programs than all finance and macro/other programs except Law and Economics and Corporate Finance. 

\begin{figure}[thbp]
  \centering
  \begin{subfigure}[t]{0.5\textwidth}
    \includegraphics*[width=\linewidth]{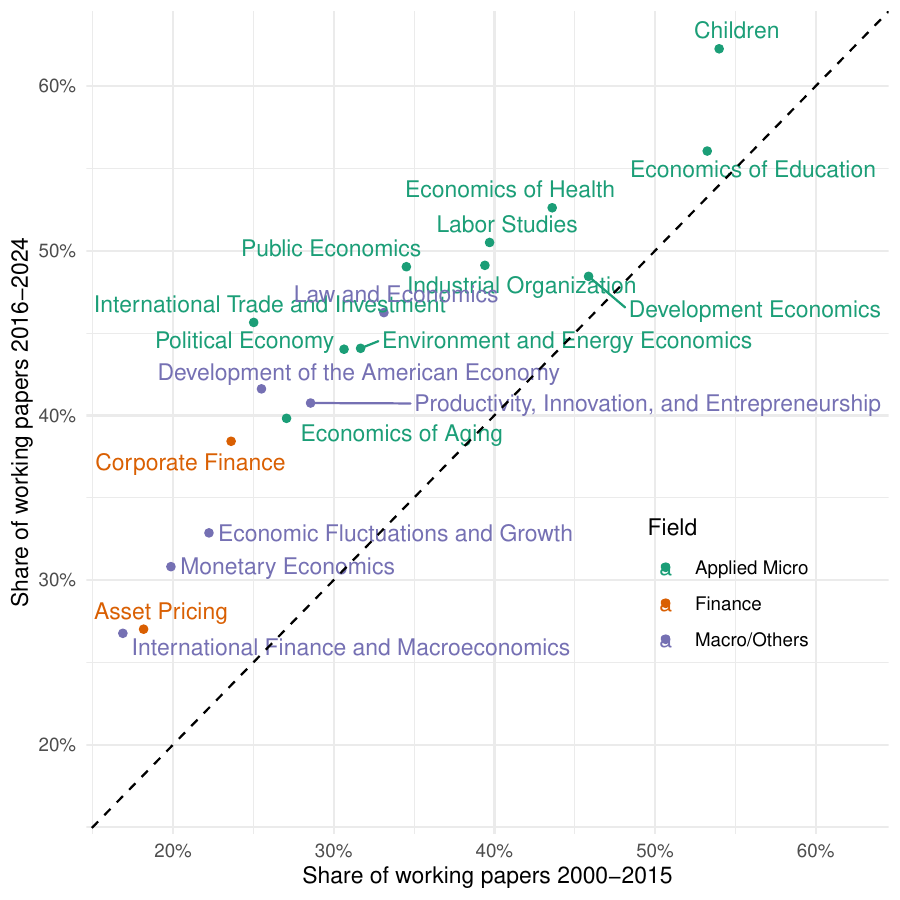}
    \caption{Identification}
    \label{fig:identification_over_time}
  \end{subfigure}%
  ~ 
  \begin{subfigure}[t]{0.5\textwidth}
    \includegraphics*[width=\linewidth]{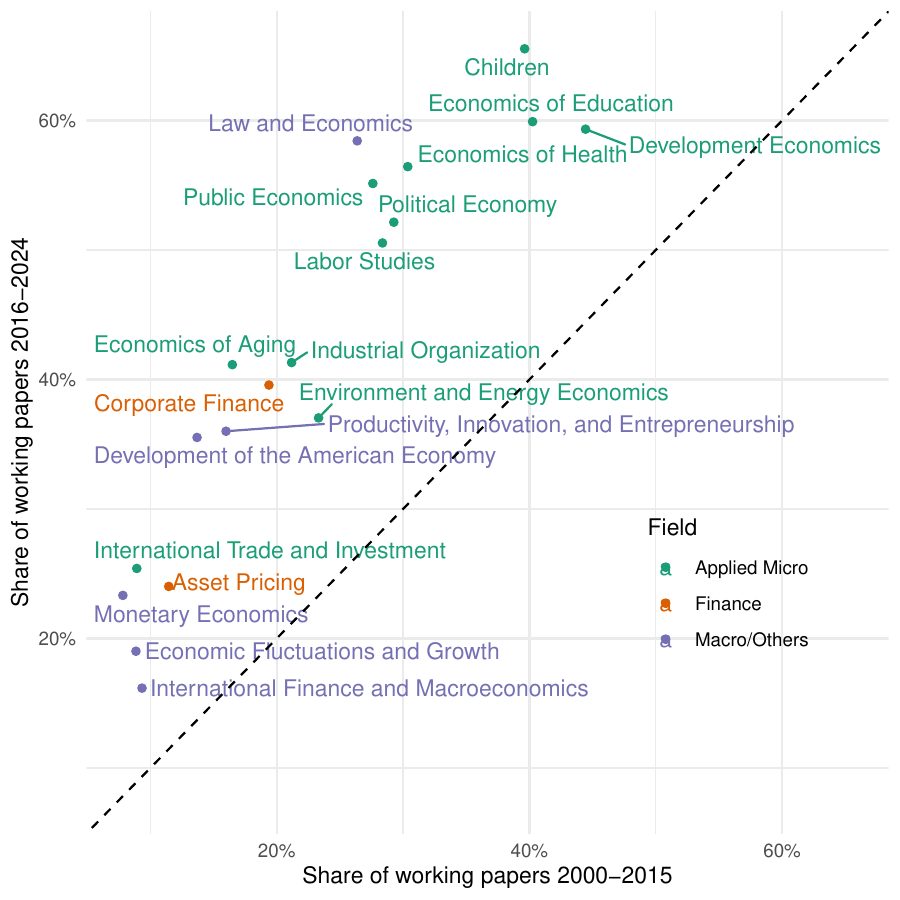}
    \caption{All experimental and quasi-experimental methods}
    \label{fig:experimentshare_over_time}
  \end{subfigure}
  \caption{This figure splits out by papers into each of the research programs for which a paper can be submitted, broken into 2000-2015 (x-axis) and 2016-2024 (y-axis). Papers can be included in more than one research program. Panel (a) reports the share of papers that mention identification strategies or concerns. Figure (b) reports the share of papers that mention any experimental or quasi-experimental method (this includes diff-in-diff, event studies, regression discontinuity, randomized control trials, lab experiments, bunching designs, and instrumental variables).  See \Cref{tab:field_groups} for the breakdown of fields, and the Appendix for definitions on keywords.}
  \label{fig:extend_bygroup_overtime}
\end{figure}

How has this changed over time for these groups? In \Cref{fig:extend_bygroup_overtime}, I split the sample into pre-2016 and 2016 and after, and examine the same shares as in \Cref{fig:extend_bygroup_overtime}. On the x-axis is the share of mentions prior to 2016, and the y-axis is the mentions in 2016 to 2024. Each point reflects a program. If there was no change in the share of mentions, the point would lie on the 45 degree line. In \Cref{fig:identification_over_time}, across all fields, there has been an increase in mentions that is similar across the board. In \Cref{fig:experimentshare_over_time}, some programs have seen much larger changes than others in their mentions of experimental and quasi-experimental methods. Notably, the change for International Finance and Macroeconomics, Economics Fluctuations and Growth, and Asset Pricing have seen less growth than Corporate Finance and Children, for example.

\begin{figure}[thbp]
  \centering
  \begin{subfigure}[t]{0.5\textwidth}
    \includegraphics*[width=\linewidth]{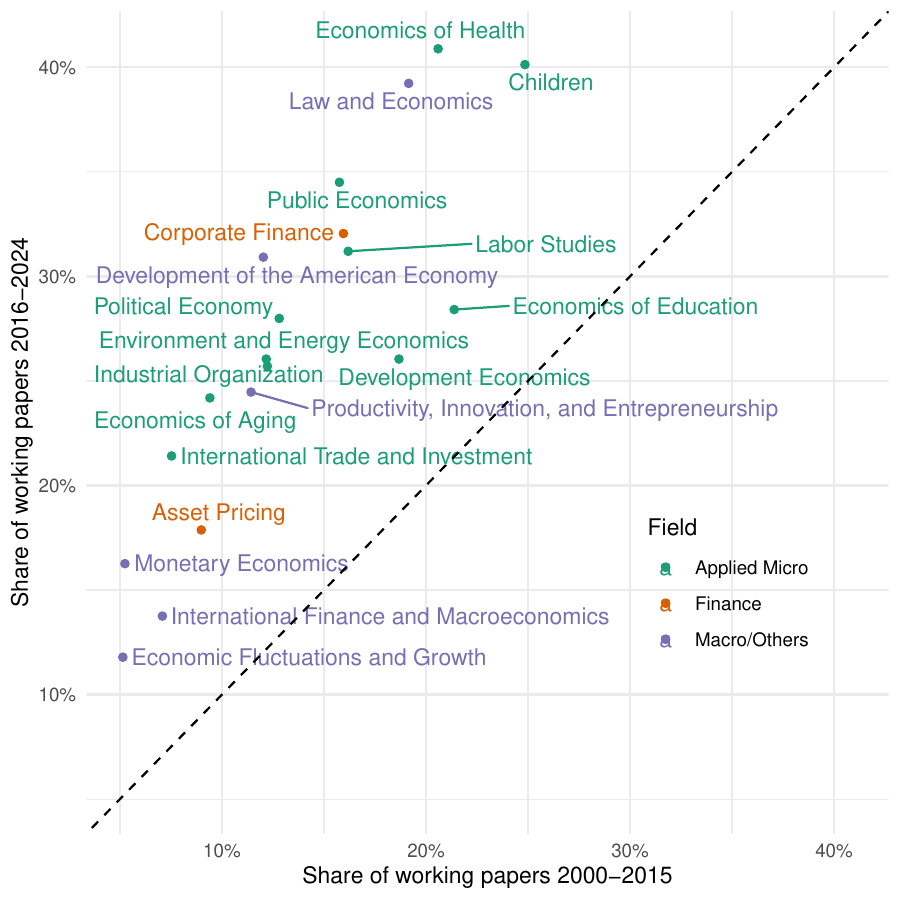}
    \caption{Difference-in-differences}
    \label{fig:did_overtime}
  \end{subfigure}%
  ~ 
  \begin{subfigure}[t]{0.5\textwidth}
    \includegraphics*[width=\linewidth]{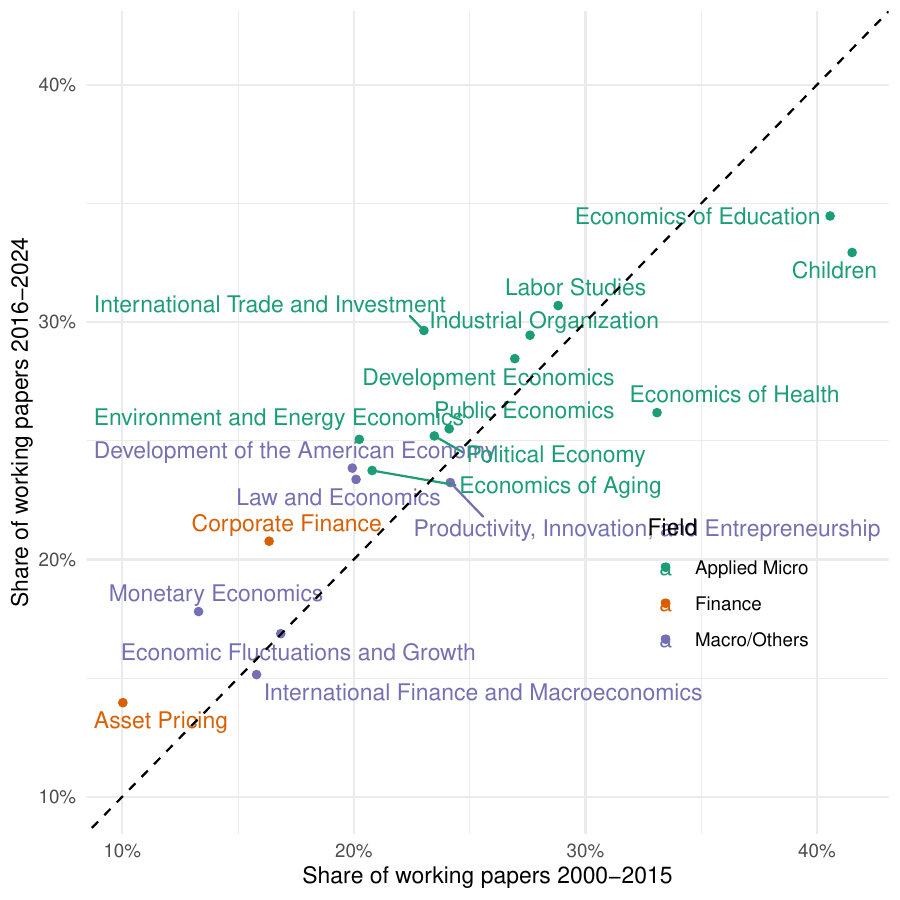}
    \caption{Instrumental variables}
    \label{fig:iv_over_time}
  \end{subfigure}
  
  \begin{subfigure}[t]{0.5\textwidth}
    \includegraphics*[width=\linewidth]{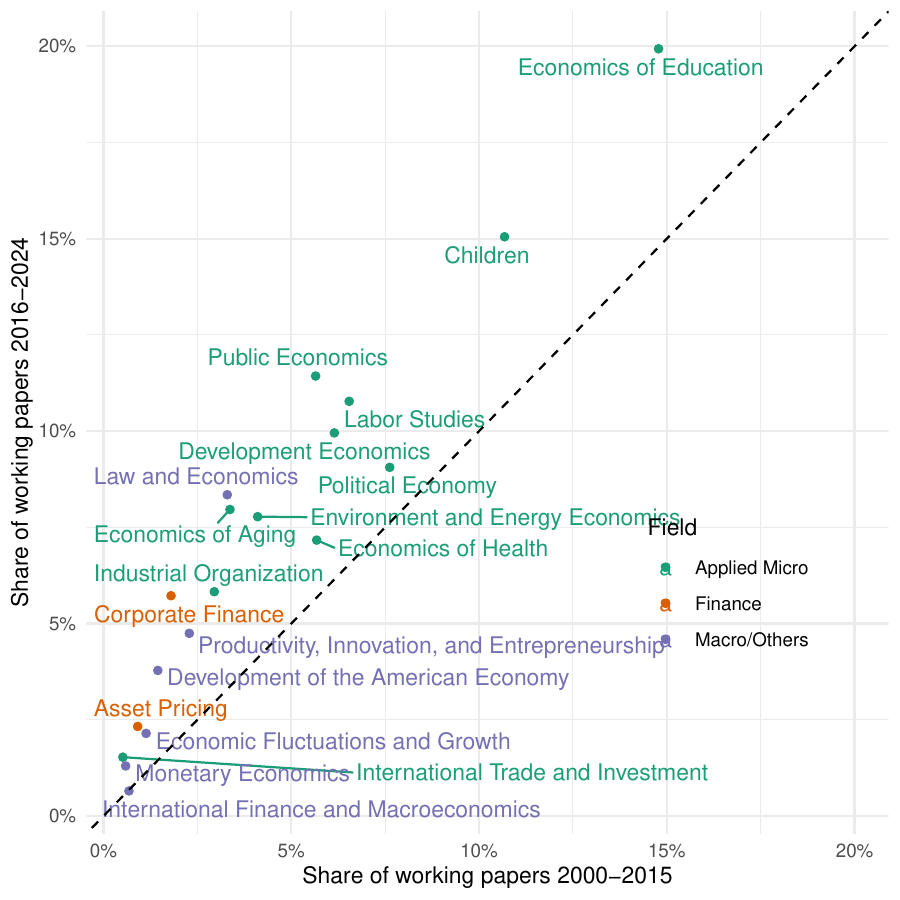}
    \caption{Regression discontinuity}
    \label{fig:rd_overtime}
  \end{subfigure}%
  ~ 
  \begin{subfigure}[t]{0.5\textwidth}
    \includegraphics*[width=\linewidth]{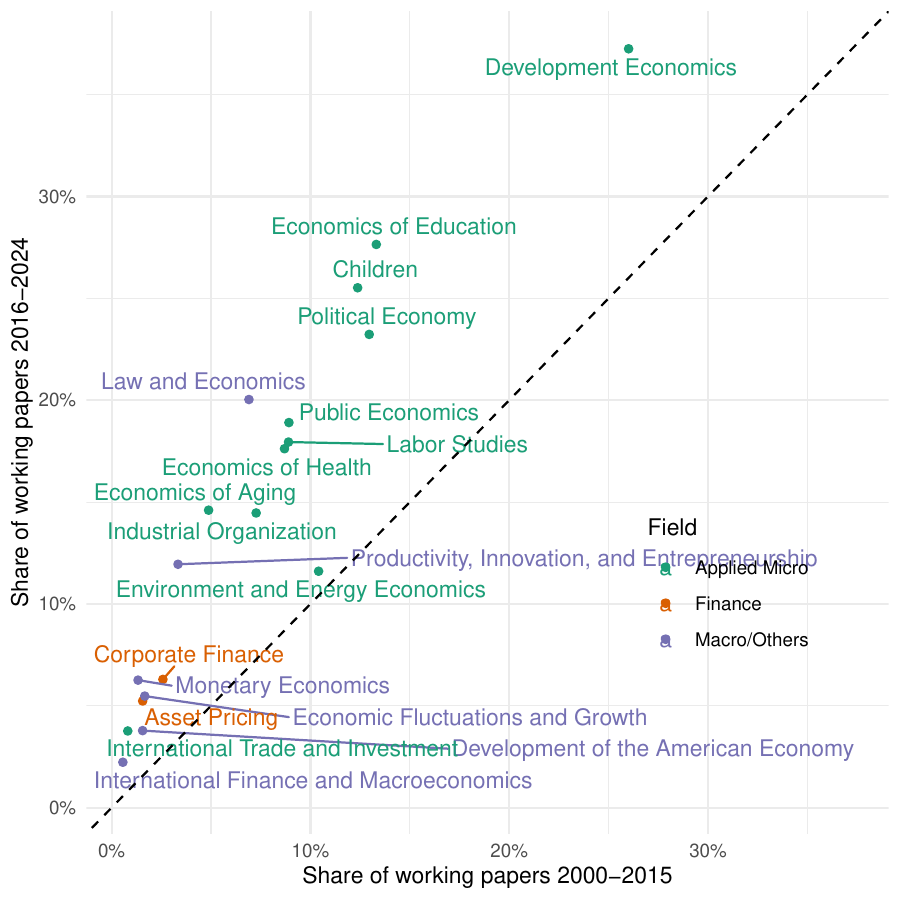}
    \caption{Experiments}
    \label{fig:experiments_overtime}
  \end{subfigure}

  \caption{This figure splits out by papers into each of the research programs for which a paper can be submitted, broken into 2000-2015 (x-axis) and 2016-2024 (y-axis). Papers can be included in more than one research program. Panel (a) reports the share of papers that mention difference-in-differences or event studies. Panel (b) reports the share of papers that mention instrumental variables. Panel (c) reports the share of papers tha mention regression discontinuity designs. Panel (d) reports the share of papers that mention RCTs or lab experiments. See \Cref{tab:field_groups} for the breakdown of fields, and the Appendix for definitions on keywords.}
  \label{fig:extend_bygroup_overtime_methods}
\end{figure}

What have been the methods that have grown significantly? In \Cref{fig:extend_bygroup_overtime_methods}, I plot the growth in difference-in-differences, instrumental variables, regression discontinuity, and experiments across programs in a similar way to \Cref{fig:extend_bygroup_overtime}. In the vast majority of progrmas, the growth has been driven by difference-in-differences, as seen in \Cref{fig:did_overtime}. In \Cref{fig:iv_over_time}, the share of papers mentioning instrumental variables has stayed roughly constant across all programs. In \Cref{fig:rd_overtime}, the share of papers mentioning regression discontinuity has risen slightly across all programs, but the growth has been more muted than difference-in-differences.\footnote{Interestingly, most of the growth and level of RD is concentrated in Public Economics, Economics of Education and Children. Education is perhaps unsurprising given the role of test scores as a canonical RD design.} Finally, in \Cref{fig:experiments_overtime}, the share of papers mentioning experiments has risen across all programs, but the growth is concentrated in a few programs (such as Development). 

Notably, for finance (and macro), there has been very limited growth in almost all methods \emph{except} difference-in-differences. This suggests that the credibility revolution in finance and macro has been driven by difference-in-differences, and not by other methods.

\begin{figure}[thbp]
  \centering
  
\begin{subfigure}[t]{0.5\textwidth}
  \includegraphics*[width=\linewidth]{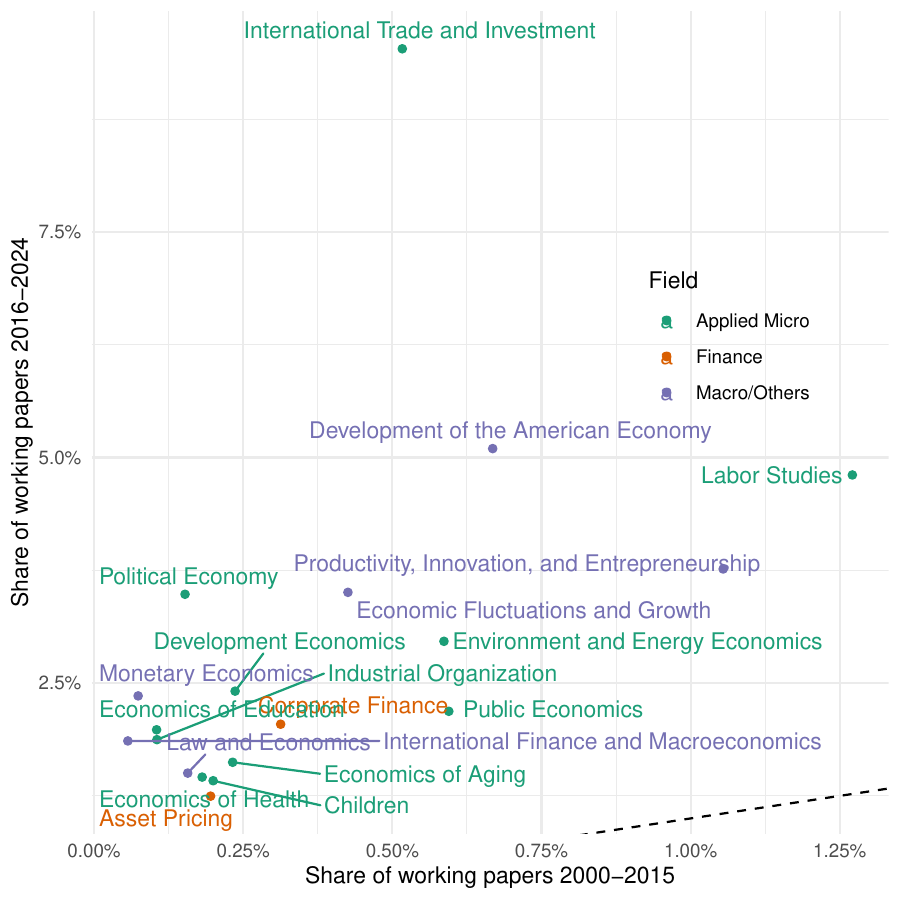}
  \caption{Bartik or shift-share instruments}
  \label{fig:bartik_overtime}
\end{subfigure}%
~ 
\begin{subfigure}[t]{0.5\textwidth}
  \includegraphics*[width=\linewidth]{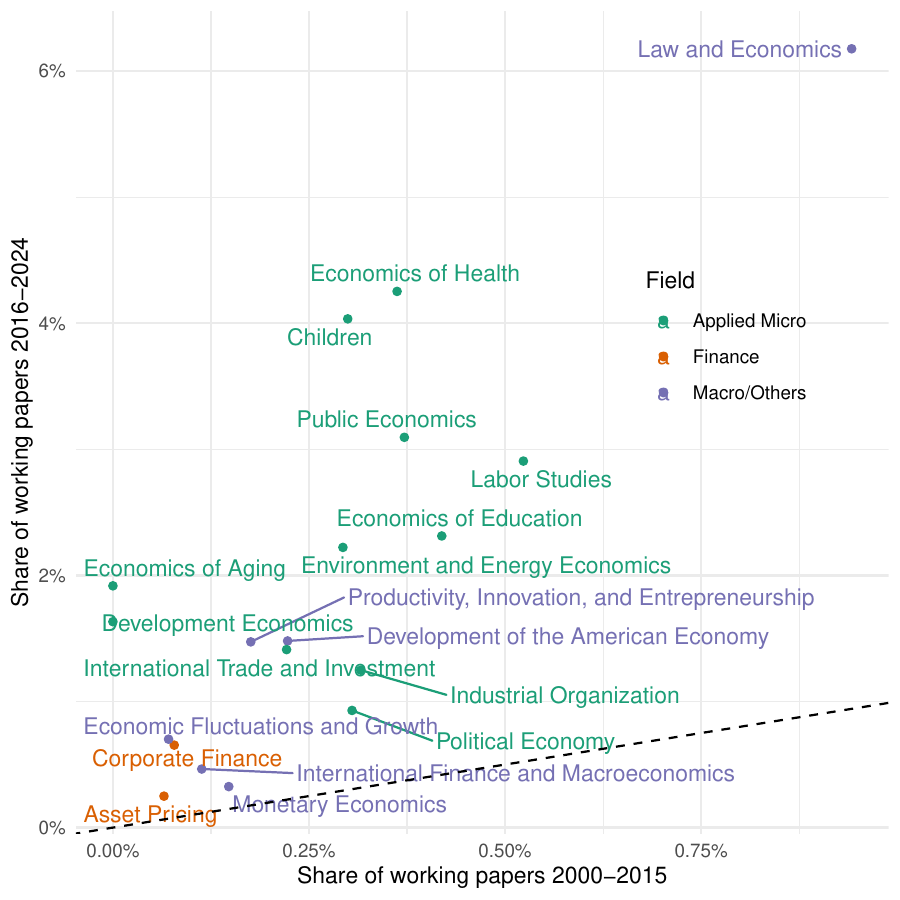}
  \caption{Synthetic controls}
  \label{fig:synth_overtime}
\end{subfigure}

  \caption{This figure splits out by papers into each of the research programs for which a paper can be submitted, broken into 2000-2015 (x-axis) and 2016-2024 (y-axis). Papers can be included in more than one research program. Panel (a) reports the share of papers that mention Bartik or shift-share instruments. Panel (b) reports the share of papers that mention synthetic control methods. See \Cref{tab:field_groups} for the breakdown of fields, and the Appendix for definitions on keywords.}
  \label{fig:bartik_synth_overtime}
\end{figure}

Lastly, I examine the tremendous recent growth of Bartik and Synthetic controls across programs in \Cref{fig:bartik_synth_overtime}. In \Cref{fig:bartik_overtime}, the share of papers mentioning Bartik and shift-share has grown across all programs, but the growth has been most pronounced in International Trade and Investment, where almost 10\% of papers mention shift-share or Bartik instruments, Development of the American Economy, and Labor Studies.\footnote{This is likely driven by the use of these instruments in studying the China Shock, popularized by \textcite{autor2013china}, the use of shift-share instruments in studying historical migration popularized by \textcite{derenoncourt2022can} in economic history, and the historical use of Bartik instruments in labor studies \parencite{bartik1991benefits}.} In \Cref{fig:synth_overtime}, the share of papers mentioning synthetic controls has grown across all programs, but the growth has been most pronounced in Law and Economics, the Economics of Health, and Children.

\begin{figure}[thbp]
  \centering
  \begin{subfigure}[t]{0.5\textwidth}
    \includegraphics*[width=\linewidth]{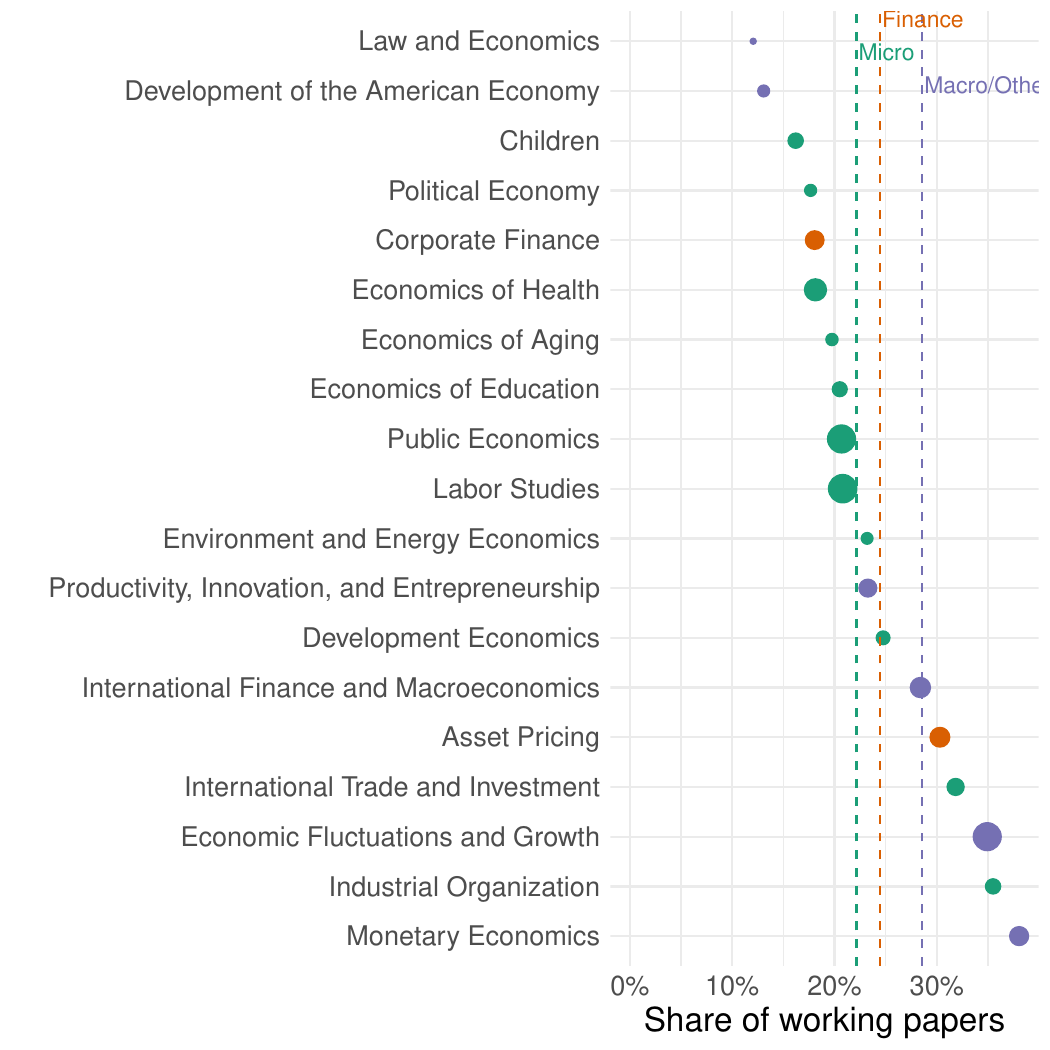}
    \caption{Stuctural estimation}
    \label{fig:structural_bygroup}
  \end{subfigure}%
  ~ 
  \begin{subfigure}[t]{0.5\textwidth}
    \includegraphics*[width=\linewidth]{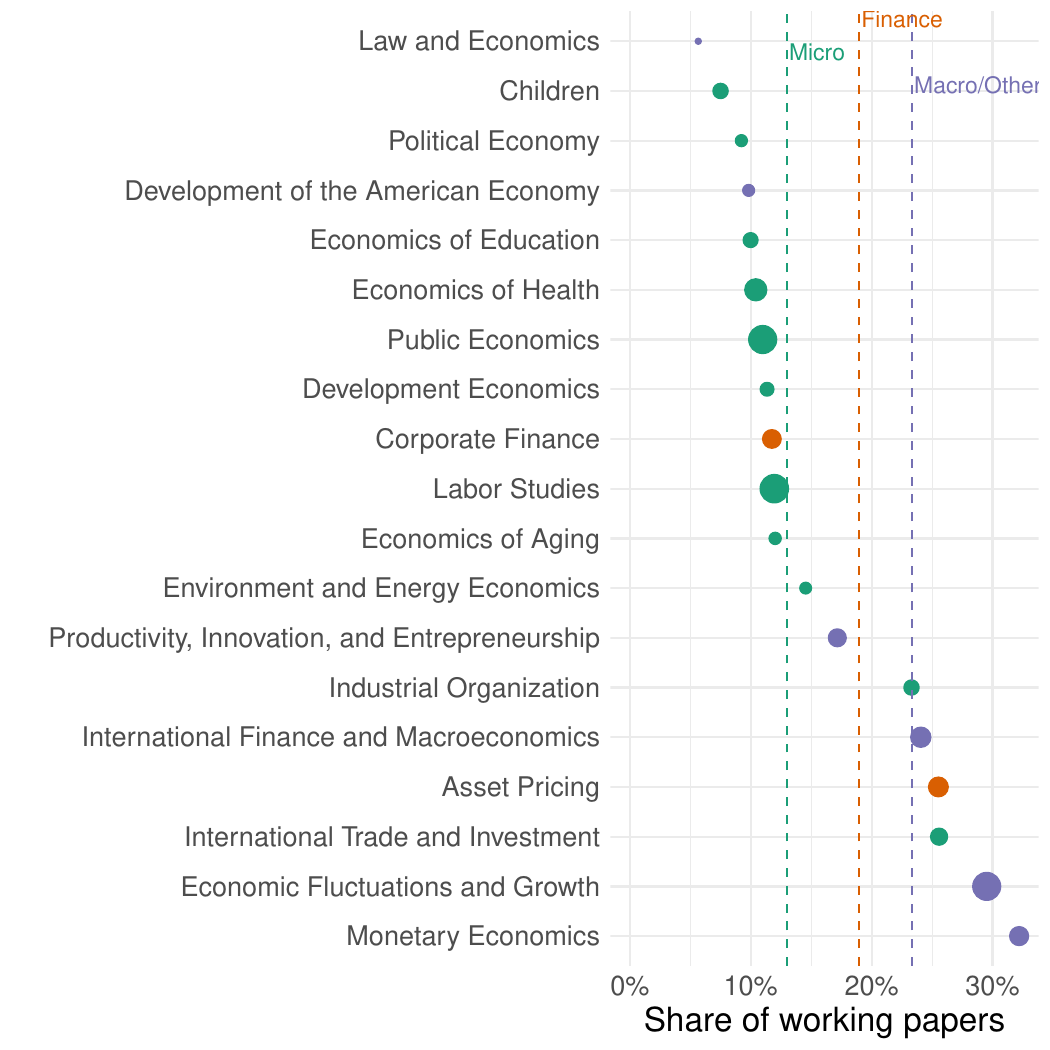}
    \caption{Structural estimation without mention of experimental and quasi-experimental methods}
    \label{fig:structural_noquasi_bygroup}
  \end{subfigure}
  \caption{This figure splits out by papers into each of the research programs for which a paper can be submitted. The size of each dot reflects the total number of papers in the program. The vertical dotted lines are the average for each field. Papers can be included in more than one research program. Panel (a) reports the share of papers that mention structural estimation. Panel (b) reports the share of papers that mention structural estimation without any mention of experimental or quasi-experimental methods (this includes diff-in-diff, event studies, regression discontinuity, randomized control trials, lab experiments, bunching designs, and instrumental variables).  See \Cref{tab:field_groups} for the breakdown of fields, and the Appendix for definitions on keywords.}
  \label{fig:struct_bygroup}
\end{figure}

How does structural estimation look across groups? In \Cref{fig:structural_bygroup}, I plot the overall share of mentions of structural estimation by program. We see a wide range of programs with shares above 30\%: Monetary Economics, Industrial Organization, EFG, International Trade and Investment, and Asset Pricing. Hence, even within each field there is tremendous heterogeneity. However, it is quite striking to examine mentions of structural models without explicit mentions of experimental and quasi-experimental methods in \Cref{fig:structural_noquasi_bygroup}. Here, we see large segmentation across programs. Most programs have 10\% or less, while some programs have 20\% or more. These include Industrial Organization, International Finance and Macroeconomics, Asset Pricing, International Trade, EFG and Monetary Economics. This suggests that in some programs, structural models are used as a primary research approach, while in others they are used as a complement to experimental and quasi-experimental methods.

\subsection{The dominance of difference-in-differences across fields}
\begin{figure}[thbp]
  \centering
  \begin{subfigure}[c]{0.5\textwidth}
    \includegraphics*[width=\linewidth]{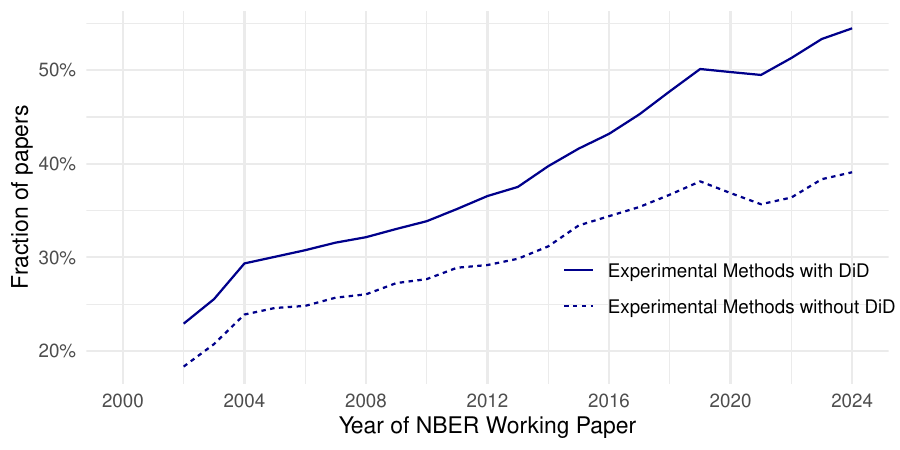}
    \caption{Effect of DiD overall}
    \label{fig:did_effect_overall}
  \end{subfigure}%
  ~ 
  \begin{subfigure}[c]{0.5\textwidth}
    \includegraphics*[width=\linewidth]{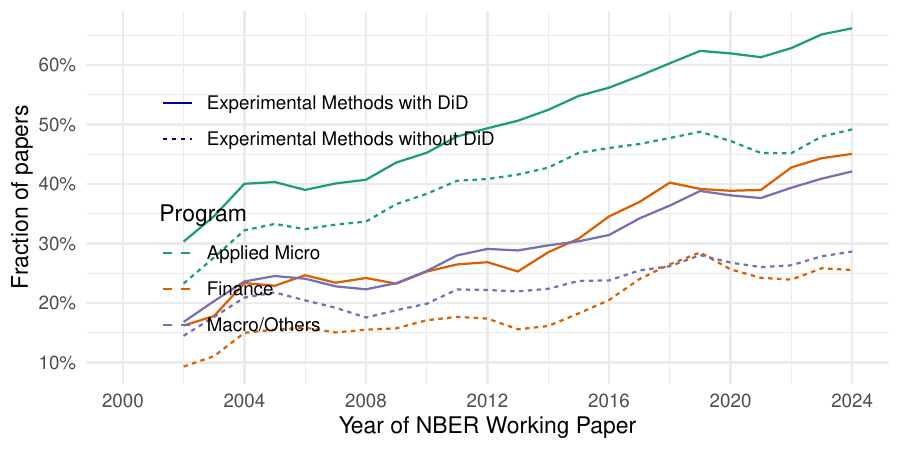}
    \caption{Effect of DiD by field}
    \label{fig:did_effect_byfield}
  \end{subfigure}%

\begin{subfigure}[c]{0.5\textwidth}
  \includegraphics*[width=\linewidth]{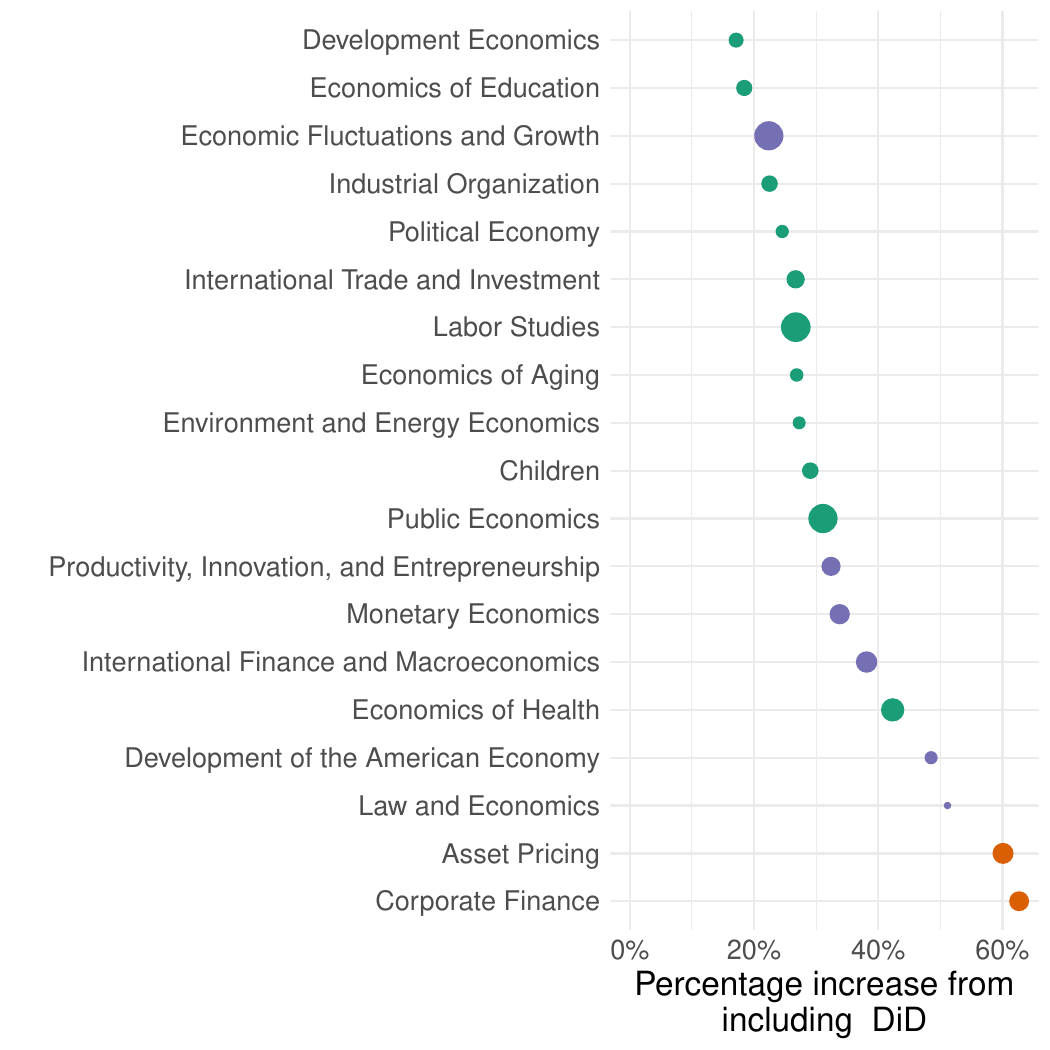}
  \caption{Percent increase from including DiD, full sample}
  \label{fig:did_effect_bygroup}
\end{subfigure}%
~ 
\begin{subfigure}[c]{0.5\textwidth}
  \includegraphics*[width=\linewidth]{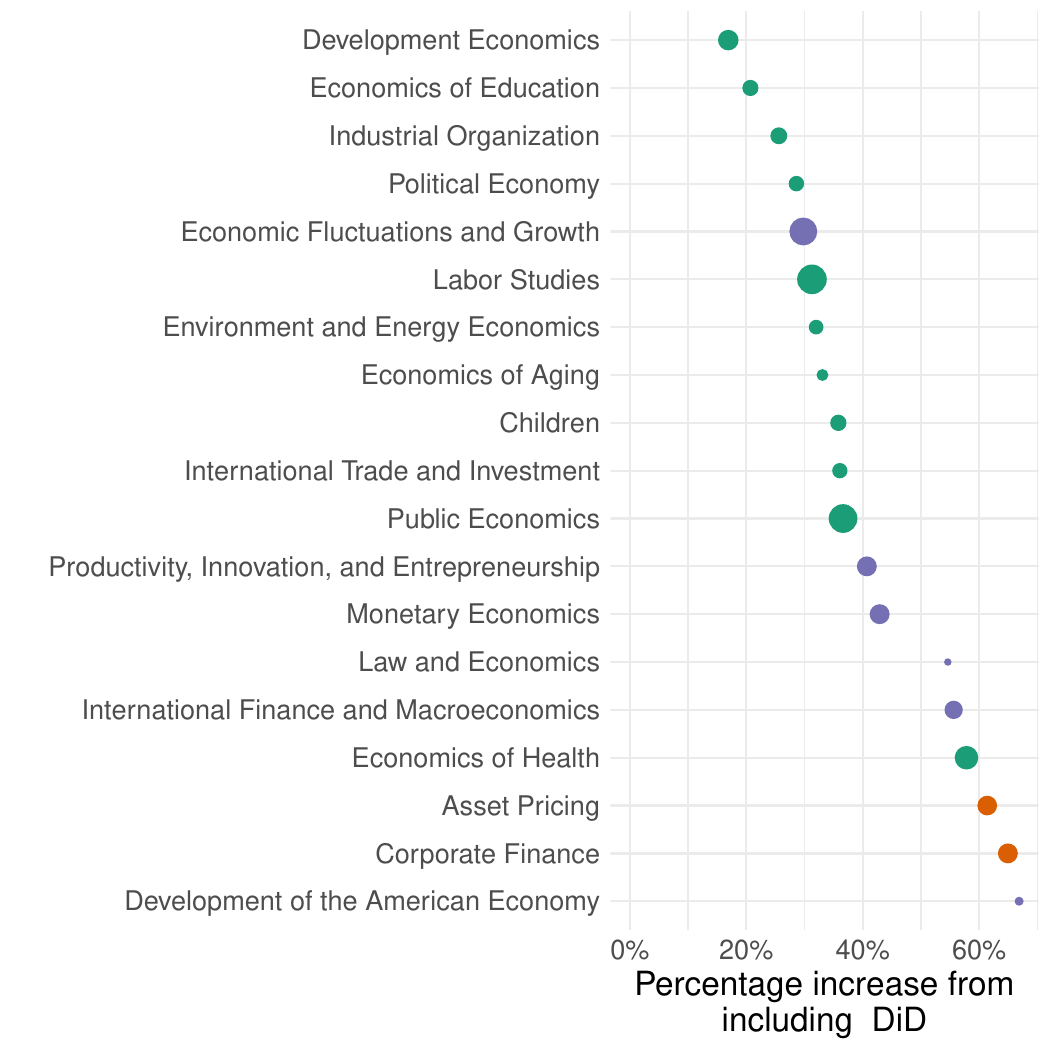}
  \caption{Percent increase from including DiD, 2016-2024}
  \label{fig:did_effect_bygroup_late}
\end{subfigure}
  \caption{This figure compares the level of mentions of experimental and quasi-experimental methods when excluding difference-in-difference methods. Panel (a) looks overall, where the solid line is all methods, as before, and the dashed line excludes any mention of DiD. Panel (b) repeats the exercise, split across fields. Panel (c) compares the percent increase in our measure of experimental and quasi-experimental mentions from including DiD, relative to not inluding the measure, over the full sample. Panel (d) repeats Panel(c), but only in the 2016-2024 period. See \Cref{tab:field_groups} for the breakdown of fields, and the Appendix for definitions on keywords.}
  \label{fig:did_effect}
\end{figure}

In the previous section, I showed that difference-in-differences has grown across all fields. It turns out that this growth in difference-in-difference has significantly increased the overall growth in experimental and non-experimental methods. In \Cref{fig:did_effect}, I document this relative impact of including DiD as a measure in the experimental and quasi-experimental term. In \Cref{fig:did_effect_overall}, I plot the overall increase in these methods, with and without DiD. As of 2024, methods are roughly 15 percentage points higher (roughly 37.5\%) than they would be without DiD. In \Cref{fig:did_effect_byfield}, I plot this change over time by field. For all fields, this gap exists, but as a percentage it is largest for finance, in part due to a lower overall level without DiD. As of 2024, there is a gap of 15 p.p. for applied micro (30\%), and 20 p.p. for finance (80\%). Hence, finance has been driven mostly heavily by DiD.

In \Cref{fig:did_effect_bygroup}, I split out by research program the average percentage increase of the methods measure from not including DiD to including DiD, over the full sample. Asset Pricing and Corporate Finance have the largest increases, as do Law and Economics and Development of the American Economy. In \Cref{fig:did_effect_bygroup_late}, I repeat the exercise but focus on the last eight years. A very similar pattern holds up for finance. In both cases, there are several applied micro fields that have high levels of mentions of experimental and quasi-experimental methods (such as development and education) that have not been affected much by difference-in-difference.

\section{Conclusion}
The credibility revolution has continued through economics over the last two decades, but there remains significant heterogeneity across fields. Applied microeconomics continues to lead the way in adopting empirical methods focused on research design and identification, with finance and macroeconomics lagging behind but still experiencing growth since the early 2000s. Notably, the growth in methods in finance and macro has been driven primarily by the adoption of difference-in-differences designs, while other quasi-experimental approaches like regression discontinuity, RCTs, and instrumental variables have seen less growth.

Looking across programs, corporate finance has seen more adoption of credibility revolution methods compared to asset pricing, explaining much of the overall growth in finance. There is also significant variation within the "macro/other" fields. Nonetheless, papers with no mention of credible empirical methods but that do mention structural estimation are more common in macro, finance, and certain applied micro fields like industrial organization, highlighting the continued importance of economic theory and structural models in many parts of the discipline.

The growing interest in and impact of difference-in-differences research across economics highlights how a single empirical technique, when widely adopted, can meaningfully shift the trajectory of an entire academic field. However, given some of the recent econometrics work flagging sensitivities and weakness in difference-in-differences, there may be value in researchers attempting to more broadly diversify their research methods portfolio \parencite{roth2022pretest, roth2023parallel, rambachan2023more, callaway2024difference, de2020two, de2022difference}. It is also quite striking that given the popularity of differnece-in-difference that synthetic control methods have not grown further, as these methods have very similar properties.

\clearpage
\printbibliography
\clearpage
\appendix
\begin{landscape}
\singlespacing
\footnotesize

\begin{longtable}{p{0.2\textwidth}p{0.5\textwidth}p{0.15\textwidth}p{0.1\textwidth}p{0.1\textwidth}p{0.1\textwidth}}
\caption{Search Categories and Trigger Phrases}\\
\hline
\textbf{Category} & \textbf{Trigger Phrases} & \textbf{Outcome} & \textbf{Case Sensitive} & \textbf{Wildcard at end} & \textbf{Cond. on 'data'} \\
\hline
\endfirsthead

\multicolumn{6}{c}%
{\tablename\ \thetable\ -- \textit{Continued from previous page}} \\
\hline
\textbf{Category} & \textbf{Trigger Phrases} & \textbf{Outcome} & \textbf{Case Sensitive} & \textbf{Wildcard at end} & \textbf{Cond. on 'data'} \\
\hline
\endhead

\hline \multicolumn{6}{r}{\textit{Continued on next page}} \\
\endfoot

\hline
\endlastfoot

Administrative Data & 'administrative data', 'admin data', 'administrative-data', 'admin-data', 'administrative record', 'admin record', administrative regist', 'admin regist', 'register data', 'registry data' & Fraction of papers with at least 1 phrase & No & Yes & Yes \\
\hline
Big Data & 'big data', 'big-data' & Fraction of papers with at least 1 phrase & No & Yes & Yes \\
\hline
Binscatter & 'binscatter', 'bin scatter', 'binned scatter' & Fraction of papers with at least 1 phrase & No & Yes & No \\
\hline 
Bunching & 'bunching' & Fraction of papers with at least 1 phrase & No & Yes & No \\
\hline
Clustering & 'cluster' & Fraction of papers with at least 1 phrase & No & Yes & Yes \\
\hline
Confidence Interval & 'confidence interval' & Fraction of papers with at least 1 phrase & No & Yes & Yes \\
\hline
Data & 'data' & Fraction of papers with at least 1 phrase & No & Yes & No \\
\hline
Difference-in-Differences & 'Difference in Diff', 'Difference in diff', 'difference in diff', 'Difference-in-Diff', 'Difference-in-diff', 'difference-in-diff', 'Differences in Diff', 'Differences in diff', 'differences in diff', 'Differences-in-Diff', 'Differences-in-diff', 'differences-in-diff', 'diff-in-diff', 'd-in-d', 'DiD' & Fraction of papers with at least 1 phrase & Yes & Yes & No \\
\hline
Event Study & 'event stud' ' event-stud' & Fraction of papers with at least 1 phrase & No & Yes & No \\
\hline
External Validity & 'external validity', 'external-validity', 'externally valid', 'externally-valid' & Fraction of papers with at least 1 phrase & No & Yes & No \\
\hline
Figure & 'graph', 'figure', 'plot', 'chart' & Average word count per paper & No & Yes & No \\
\hline
Fixed Effects & 'FE', 'Fixed Effect', 'Fixed effect', 'fixed effect', Fixed Effects', 'Fixed effects', 'fixed effects', 'Fixed-Effect', 'Fixed-effect', 'fixed-effect', 'Fixed-Effects', Fixed-effects', 'fixed-effects' & Fraction of papers with at least 1 phrase & Yes & No & Yes \\
\hline
Functional Forms & 'CES', 'constant elasticity of substitution', 'Constant Elasticity of Substitution', 'Constant elasticity of substitution', 'Cobb-Douglas', 'Cobb Douglas', 'Stone Geary', 'Stone-Geary', 'CRRA', 'coefficient of relative risk-aversion', 'coefficient of relative risk aversion', 'Coefficient of relative risk-aversion', 'Coefficient of relative risk aversion', 'Coefficient of Relative Risk-Aversion', 'Coefficient of Relative Risk Aversion', 'CARA', 'constant absolute risk aversion', 'constant absolute risk-aversion', 'Constant absolute risk aversion', 'Constant absolute risk-aversion', 'Constant Absolute Risk Aversion', 'Constant Absolute Risk-Aversion', 'translog', 'Translog' & Fraction of papers with at least 1 phrase & Yes & No & No \\
\hline
General Equilibrium & 'general equilibr', 'general-equilibr' & Fraction of papers with at least 1 phrase & No & Yes & No \\
\hline
Identification & Sentence structure: search for sentences that have the term 'identif' in combination with any of the terms: 'effect', 'response', 'impact', 'elasticit', 'parameter', or 'coefficient' with maximum two words in between. Note that even though the search includes wildcards at the end, we exclude any match with the word 'effective'. Also search for these terms: 'causal identification', 'causally identified', 'identification strategy', 'identification approach', 'identification assumption', 'identifying assumption', 'identifying variation', 'empirical identification', 'over identified', 'over-identified', 'under identified', 'under-identified', 'identification properties', 'identification test', 'identification problem', 'identification issue', 'problem with identification', 'problems with identification', 'issue with identification', 'issues with identification', 'problem identifying', 'problems identifying', 'issue identifying','issues identifying', 'threat to identification', 'threats to identification', 'threat for identification', 'threats for identification', 'over identifying', 'over-identifying', 'under identifying', 'under-identifying', 'partial identification', 'partially identified', 'non-parametric identification', 'nonparametric identification', 'non parametric identification', 'non-parametrically identified', 'nonparametrically identified', 'non parametrically identified', 'identification condition', 'identifying condition', 'condition for identification', 'conditions for identification', 'condition for identifying', 'conditions for identifying', 'point identification', 'point-identification', 'point identified', 'point-identified', 'point identifying', 'point-identifying', 'set identification', 'set-identification', 'set identified', 'set-identified', 'set identifying', 'set-identifying', 'identification analysis', 'weak identification', 'identification result', 'identification argument', 'identification framework', 'identification scheme' & Fraction of papers with at least 1 phrase & No & Yes & No \\
\hline
Internet Data & 'internet data', 'internet-data', 'web data', 'web-data', 'scraped data', 'scraped-data', 'scrape data', 'scraping data', 'search data', 'search-data', 'google data', google-data', 'social media data', 'google trend', 'google-trend', 'google search', 'google-search', 'google ngram', 'google n-gram', 'google books ngram', 'google books n-gram' & Fraction of papers with at least 1 phrase & No & Yes & Yes \\
\hline
Instrumental Variables & 'Instrumental Variable', 'Instrumental variable', 'instrumental variable', 'Instrumental-Variable', 'Instrumental-variable', 'instrumental-variable', 'Two Stage Least Squares', 'Two stage least squares', 'two stage least squares', '2SLS', 'TSLS', 'valid instrument', 'exogenous instrument', 'IV Estimat', 'IV estimat', 'IV-estimat', 'IV Specification', 'IV specification', 'IV-specification', 'IV Regression', 'IV regression', 'IV-regression', 'IV Strateg', 'IV strateg', 'IV-strateg', 'we instrument', 'I instrument', 'paper instruments', 'exclusion restriction', 'weak first stage', 'simulated instrument' & Fraction of papers with at least 1 phrase & Yes & Yes & Yes \\
\hline
Lab Experiments & 'Laboratory Experiment', 'Laboratory experiment', 'laboratory experiment', 'Lab Experiment', 'Lab experiment', 'lab experiment', 'Dictator Game', 'Dictator game', 'dictator game', 'Ultimatum Game', 'Ultimatum game', 'ultimatum game', 'Trust Game', 'Trust game', 'trust game' , 'Public Good Game', 'Public good game', 'public good game', 'Public Goods Game', 'Public goods game', 'public goods game', 'Z-tree', 'zTree', 'ORSEE', 'show-up fee', 'laboratory participant', 'lab participant' & Fraction of papers with at least 1 phrase & Yes & Yes & No \\
\hline
Machine Learning & 'machine learning', 'lasso', 'random forest' & Fraction of papers with at least 1 phrase & No & Yes & No \\
\hline
Matching & 'propensity score', 'propensity score matching', 'propensity-score matching', 'matching estimat', 'nearest neighbor matching', 'nearest-neighbor matching', 'nearest neighbour matching', 'nearest-neighbour matching', 'caliper matching', 'stratification matching', 'exact matching', 'one to one matching', 'one-to-one matching', 'kernel matching', 'inverse probability matching', 'inverse-probability matching' & Fraction of papers with at least 1 phrase & No & Yes & Yes \\
\hline
Mechanisms & 'mechanism' & Fraction of papers with at least 1 phrase & No & Yes & No \\
\hline
Omitted Variables & 'omitted variable' & Fraction of papers with at least 1 phrase & No & Yes & Yes \\
\hline
Preanalysis Plan & 'pre-analysis plan', 'pre analysis plan', 'preanalysis plan' & Fraction of papers with at least 1 phrase & No & Yes & No \\
\hline
Precisely Estimated & 'precisely estimated', 'precisely-estimated' & Fraction of papers with at least 1 phrase & No & Yes & No \\
\hline
Precisely Estimated Zero & 'precisely estimated zero', 'precisely-estimated zero' & Fraction of papers with at least 1 phrase & No & Yes & No \\
\hline
Proprietary Data & 'proprietary data', 'confidential data', 'nonpublic data', 'non-public data', 'proprietary-data', 'confidential-data', 'nonpublic-data', 'non-public-data' & Fraction of papers with at least 1 phrase & No & Yes & Yes \\
\hline
Quasi- and Natural Experiments & 'quasi experiment', 'quasi-experiment', 'quasiexperiment', 'natural experiment', 'natural-experiment' & Fraction of papers with at least 1 phrase & No & Yes & No \\
\hline
RCTs & 'Randomized Controlled Trial' , 'Randomized controlled trial', 'randomized controlled trial', 'Randomized Control Trial', 'Randomized control trial', 'randomized control trial', 'Randomized Field Experiment', 'Randomized field experiment', 'randomized field experiment', 'Randomized Controlled Experiment', 'Randomized controlled experiment', 'randomized controlled experiment', 'Randomised Controlled Trial' , 'Randomised controlled trial', 'randomised controlled trial', 'Randomised Control Trial', 'Randomised control trial', 'randomised control trial', 'Randomised Field Experiment', 'Randomised field experiment', 'randomised field experiment', 'Randomised Controlled Experiment', 'Randomised controlled experiment', 'randomised controlled experiment', 'Social Experiment', 'Social experiment', 'social experiment', 'RCT' & Fraction of papers with at least 1 phrase & Yes & Yes & No \\
\hline
Regression Discontinuity & 'Regression Discontinuit', 'Regression discontinuit', 'regression discontinuit', 'Regression-discontinuity', 'regression-discontinuity', 'Regression Kink', 'Regression kink', 'regression kink', 'RD Design', 'RD design', RD-design', 'RD Estimat', 'RD estimat', 'RD-estimat', 'RD Model', 'RD model', 'RD-model' , 'RD Regression', 'RD regression', 'RD-regression', 'RD Coefficient', 'RD coefficient', 'RD-coefficient', 'RK Design', 'RK design', 'RK-Design', 'RK-design', 'RKD', 'RDD' & Fraction of papers with at least 1 phrase & Yes & Yes & No \\
\hline
Reverse Causation & 'reverse causa', 'reverse-causa' & Fraction of papers with at least 1 phrase & No & Yes & Yes \\
\hline
Selection & 'selection' & Fraction of papers with at least 1 phrase & No & Yes & Yes \\
\hline
Simultaneity & 'simultaneity' & Fraction of papers with at least 1 phrase & No & Yes & Yes \\
\hline
Structural Model & Sentence structure: we search for instances where, within two full stops, the term 'structural' is mentioned in combination with either 'model', 'specification', 'estimate', or 'parameter'. Also search for these terms: 'Structural Model', 'Structural model', 'structural model', 'Method of Moments', 'Method of moments', 'method of moments', 'Method-of-Moments', 'Method-of-moments', 'method-of-moments', 'Berry, Levinsohn, Pakes', 'Berry, Levinsohn and Pakes', 'Berry, Levinsohn, and Pakes', 'BLP', 'Structural General Equilibrium Model', 'Structural general equilibrium model', 'structural general equilibrium model', 'GMM', 'Maximum Likelihood Estimat', 'Maximum likelihood estimat', 'maximum likelihood estimat', 'Maximum-Likelihood Estimat', 'Maximum-likelihood estimat', 'maximum-likelihood estimat', 'MLE' & Fraction of papers with at least 1 phrase & Yes & Yes & No \\
\hline
Survey Data & Sentence structure: we search for instances where the term 'survey' and 'data' are mentioned within two full stops. & Fraction of papers with at least 1 phrase & No & Yes & Yes \\
\hline
Synthetic Control & 'synthetic control' & Fraction of papers with at least 1 phrase & No & Yes & Yes \\
\hline
Table & 'table' & Average word count per paper & No & Yes & No \\
\hline
Text Analysis & 'natural language processing', 'text analys', 'computational linguistics', 'speech processing', 'n-gram', 'ngram', 'n gram', 'textual analys', 'language processing', 'language analys', 'text data', 'text mining', 'mining text', 'text regression', 'tokeniz' & Fraction of papers with at least 1 phrase & No & Yes & No \\

\end{longtable}
\end{landscape}
\end{document}